\documentclass[twocolumn,showpacs,preprintnumbers,amsmath,amssymb,superscriptaddress,aps]{revtex4}

\usepackage{graphicx}
\usepackage{dcolumn}
\usepackage{bm}

\begin{document}


\title{Measurement of the Transverse Polarization of Electrons
\\ Emitted in Free Neutron Decay}

 \address{Marian Smoluchowski Institute of Physics, Jagiellonian University, Cracow, Poland}
 \address{Henryk Niewodnicza\'nski Institute of Nuclear Physics PAN, Cracow, Poland}
 \address{Paul Scherrer Institut, Villigen, Switzerland}
 \address{LPC-Caen, ENSICAEN, Universit\'e de Caen Basse-Normandie, CNRS/IN2P3-ENSI, Caen, France}
 \address{Katholieke Universiteit Leuven, Leuven, Belgium}

\author{A. Kozela}
 \address{Henryk Niewodnicza\'nski Institute of Nuclear Physics PAN, Cracow, Poland}

\author{G. Ban}
 \address{LPC-Caen, ENSICAEN, Universit\'e de Caen Basse-Normandie, CNRS/IN2P3-ENSI, Caen, France}
\author{A. Bia{\l}ek}
 \address{Henryk Niewodnicza\'nski Institute of Nuclear Physics PAN, Cracow, Poland}

\author{K. Bodek}
 \address{Marian Smoluchowski Institute of Physics, Jagiellonian University, Cracow, Poland}
\author{P.~Gorel}
 \address{Marian Smoluchowski Institute of Physics, Jagiellonian University, Cracow, Poland}
 \address{Paul Scherrer Institut, Villigen, Switzerland}
\address{LPC-Caen, ENSICAEN, Universit\'e de Caen Basse-Normandie, CNRS/IN2P3-ENSI, Caen, France}
\author{K.~Kirch}
 \address{Paul Scherrer Institut, Villigen, Switzerland}
 \address{Eidgen\"ossische Technische Hochschule, Z\"urich, Switzerland}
\author{St.~Kistryn}
 \address{Marian Smoluchowski Institute of Physics, Jagiellonian University, Cracow, Poland}
\author{O.~Naviliat-Cuncic}
 \address{LPC-Caen, ENSICAEN, Universit\'e de Caen Basse-Normandie, CNRS/IN2P3-ENSI, Caen, France}
 \address{NSCL and Departament of Physics and Astronomy, MSU, East Lansing, MI, USA}
\author{N.~Severijns}
 \address{Katholieke Universiteit Leuven, Leuven, Belgium}
\author{E. Stephan}
 \address{Institute of Physics, University of Silesia, Katowice, Poland}
\author{J.~Zejma}
 \address{Marian Smoluchowski Institute of Physics, Jagiellonian University, Cracow, Poland}

\date{\today}
\collaboration{The nTRV Collaboration}\noaffiliation

\begin{abstract}
The final analysis of the experiment determining
both components of the transverse polarization of electrons ($\sigma_{T_{1}}$, $\sigma_{T_{2}}$) 
emitted in the $\beta$-decay of polarized, free neutrons is presented. 
The T-odd, P-odd correlation coefficient quantifying 
$\sigma_{T_{2}}$, perpendicular to the neutron polarization and 
electron momentum, was found to be $R=$ 0.004$\pm$0.012$\pm$0.005.  
This value is consistent with time reversal invariance, and significantly 
improves both earlier result and limits on the relative strength of imaginary
 scalar couplings in the weak interaction.
The value obtained for the correlation coefficient associated with $\sigma_{T_{1}}$, 
$N=$ 0.067$\pm$0.011$\pm$0.004, agrees  with 
the Standard Model expectation, providing an important sensitivity test 
of the experimental setup.
The present result sets constraints on the imaginary part of scalar and tensor couplings in weak 
interaction. 
Implications for parameters of the leptoquark exchange model and minimal supersymmetric  
model (MSSM) with $R$-parity violation are discussed.
\end{abstract}

\maketitle

\section{Introduction}

In the Standard Model (SM) the description of free neutron decay involves only three 
parameters: (i) the relative strength of axial and vector couplings, $\lambda=g_{A}/g_{V}$, 
(ii) the first element of the quark mixing matrix, $V_{ud}$, and (iii) a time reversal 
violating phase $\Phi$. 
The much larger number of observables which became accessible in novel experiments at new 
generation neutron sources allows not only to contribute to the determination 
of those parameters, but also to address some basic problems reaching 
beyond the SM.

One of these is the incomplete knowledge of the physics of  CP violation  
(combined charge conjugation and parity symmetry).
The SM with the Cabbibo-Kobayashi-Maskawa (CKM) mixing scheme \cite{Kob73} accounts for 
CP violation discovered in kaon \cite{Chris64} and B-meson \cite{Abe02,Aube02} systems.
It fails by many orders of magnitude to account for the most striking evidence of CP 
violation: the dominance of baryonic matter in the present universe \cite{Sak67, Tro99}.

Supported by firm theoretical considerations and  strong experimental evidence 
\cite{Gib93}, the combined CPT symmetry is regarded to be a strict symmetry of nature. 
With this assumption CP violation is equivalent to time reversal symmetry  
violation (TRV) and as a result is linked to microscopic reversibility and the principle of 
detailed balance.  
No compelling evidence of TRV has been observed in  experiments testing this principle in 
different nuclear reactions \cite{Dra94}, and the T-violating 
amplitude was found to be at most 10$^{-3}$ of the dominant strong interaction amplitude. 

To the most precise tests of time reversal invariance belong searches  for electric dipole 
moments of elementary particles, atoms and molecules. 
Despite their impressive accuracy  one obtains only upper bounds ($2.9 \times 10^{-26}$, 
$5.9 \times 10^{-28}$, $3.1 \times 10^{-29}$ e cm for 
neutrons \cite{Bak06}, electrons \cite{Hud11} and $^{199}$Hg atoms  \cite{Gri09}, 
respectively) which are still orders of magnitude away from the SM predictions (e.g. $10^{-32}$
--  $10^{-34}$ e cm for the neutron \cite{Czar97,Dow02}), leaving a lot of room for new physics 
searches.

The situation is more complicated in high energy experiments or in systems with heavy quarks 
contents.
Also here it is possible to construct observables sensitive to TRV, but the sizable 
contributions of heavy quarks makes it difficult to disentangle between new physics and the SM 
induced effects.
This was the case in the first direct observation of TRV in the kaon system in the CPLEAR  
\cite{Ange98} and KTeV experiments \cite{Ala00}. 
Only very recently the D0 collaboration reported the observation of a charge asymmetry  like-sign 
dimuon production in proton antiproton collisions at 1.96 TeV center of mass energy, which 
contradicts the SM at 3.2 standard deviations \cite{Aba10}. 

More than 50 years ago it was recognized that TRV may be tested also in various 
correlations accessible in nuclear or particles decays \cite{Jack57,Cur57}.  
Many systems have been investigated in this way including mesons \cite{Abe04}, leptons
\cite{Yamaz10}, baryons, and nuclei (see \cite{Sever06} for a review). 
The measurement of the $R$ coefficient in $^{8}$Li decay, quantifying the correlation between 
the spin of the decaying nucleus, the electron momentum and the electron spin, provides the 
most stringent direct limit on the imaginary part of tensor coupling constants of the weak 
interaction \cite{Hub03}. 
The discovery of new CP- or T-violating phenomena, especially in systems built 
of  quarks of the first generation, with vanishingly small contributions  
from the CKM matrix induced mechanism, would be an important milestone.
 
Free neutron decay plays a particular role in nuclear beta decay experiments searching for 
TRV. 
Due to its simplicity it is free from model dependent corrections associated with 
nuclear and atomic structure. 
Further, final state interaction effects, which can mimic T violation, are small in this case 
and can, in addition, be calculated with a relative precision better than 1\% \cite{Vog83}. 
From a variety of correlation coefficients which may be built from vectors accessible in neutron 
decay, up to now only two have been addressed experimentally. 
First was the angular correlation between the neutron spin, the electron momentum and the 
neutrino momentum, referred to as $D$ coefficient  in the literature. 
It is sensitive to the relative strength ($g_{A}/g_{V}$) and phase angle ($\varPhi_{V\!A}$) 
between axial and vector currents in weak interaction, and has been measured in several 
experiments \cite{Cla58,Lis00,Sol04,Mum11}.  
At the current precision it provides the best limits to certain time reversal violating parameters 
appearing in standard model extensions with leptoquarks exchange, associated with a non-zero 
value of $\sin{\varPhi_{V\!A}}$.  

In this paper we present the final analysis of the first measurement of another time 
reversal violating correlation coefficient in neutron decay -- the $R$ coefficient, and 
of the time reversal conserving $N$ correlation, both 
associated with a correlation between the neutron spin, the electron momentum and its polarization. 
Being sensitive to the real and imaginary parts of scalar and tensor couplings of the weak 
interaction they provide information complementary to the $D$ coefficient. 

This is  the final report of the nTRV experiment comprising data collected between 2004 and 2007. 
It supersedes our previous result, presenting the methods used in the data analysis in 
more details and introducing a new $R$-evaluation approach based on a ``double'' ratio method. 
The significant improvement in the accuracy of the determination of this  
correlation coefficient as compared to the result presented in \cite{Koz09} is
a consequence of two major extensions in the analysis of the existing data: 
 (i) the analysis of an additional event class with 
backscattered electrons trajectories contained within the vertical plane, and 
(ii) improved determination of the effective analyzing powers  of the applied Mott scatterers.
Minor changes in the value of the P-even, T-even $N$ correlation coefficient are the result of 
new effective analyzing powers and the analysis of another event class which has also not 
been included in the previous analysis.

The outline of this paper is as follows. 
In Sec.\ \ref{S2} we introduce the measured observables and present the strategy of the 
experiment.
The dedicated neutron beam line, detector setup and performance as well as the hardware 
trigger are discussed in Sec.\ \ref{S3}.
Sec.\ \ref{S4} presents the data analysis, discussion of systematic effects and describes 
applied consistency checks. 
The results obtained and their implications on some extensions of the SM are compared with 
existing 
experimental data in Sec.\ \ref{S5}, and finally, conclusions are given in Sec.\ \ref{S6}.

\section{\label{S2}  Correlations in neutron $\beta$-decay}

\begin{figure}[htb]
\includegraphics[scale=.99]{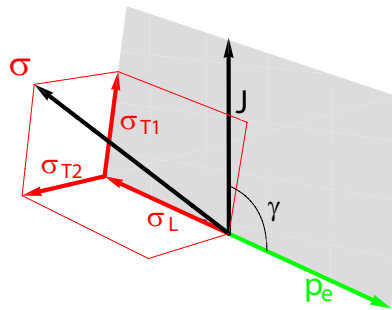}
\caption{\label{nTRV_Vectors_decay} (Color online) Schematic representation of  neutron 
decay. The decay plane containing the neutron polarization $J$, the electron momentum $p_{e}$ 
 and the transverse component of the electron polarization  $\sigma_{T_{1}}$ is indicated.}
\end{figure}

The electron distribution function for an experiment, in which the 
decaying neutrons are oriented, and electron energy, momentum ($E, {\bf p}$) and polarization 
are measured, is proportional to \cite{Jack57}: 
\begin{eqnarray}\label{Wprob}
  \nonumber
  W(\langle{\bf J}\rangle, \mbox{\boldmath$ \hat{\sigma}$}, E, {\bf p}) \varpropto
  \dfrac{}{} 1  + b\,\frac{m}{E} +\frac{{\bf p}}{E}\cdot\left(A \frac{\langle{\bf J}\rangle}{J} + G\, \mbox{\boldmath$ \hat{\sigma}$} \right) +\\
  +\frac{\langle{\bf J}\rangle}{J} \cdot \left( Q \frac{{\bf p}}{E} \frac{{\bf p\cdot\mbox{\boldmath$ \hat{\sigma}$}}}{E+m} + N {\mbox{\boldmath$ \hat{\sigma}$}} +R \frac{{\bf p} \times \mbox{\boldmath$ \hat{\sigma}$}}{E} \right) ,
\end{eqnarray}
where $m$ is the electron mass, $J$ and  $\langle{\bf J}\rangle$ are the neutron spin and its 
expectation value, respectively. 
\mbox{\boldmath$\hat{\sigma}$} is a unit vector onto 
which the electron spin is projected and $A$ is the $\beta$-decay asymmetry parameter. 
Higher order terms in ${\bf p}$, ${\bf J}$ and $\mbox{\boldmath$\hat{\sigma}$}$ are neglected. 
$N$ and $R$ are the correlation coefficients  associated with $\sigma_{T_{1}}$ and 
$\sigma_{T_{2}}$, respectively, where  $\sigma_{T_{2}}$ is the transverse component of the 
electron polarization perpendicular to the decay plane spanned by the neutron spin and 
the electron momentum, and  $\sigma_{T_{1}}$ is the component contained within this plane 
(Fig.~\ref{nTRV_Vectors_decay}).
As the terms involving $G$ and $Q$ are proportional to the longitudinal component of electron 
polarization which was not accessible in the presented experiment 
Eq.~(\ref{Wprob}) reduces to:
 \begin{align}\label{Wprob2}
\nonumber
  W({\bf J}, \mbox{\boldmath$ \hat{\sigma}$}, E, {\bf p}) &\varpropto
  \dfrac{}{} 1 + b\,\frac{m}{E} \\  
&+ {\bf P} \cdot   \left(A\frac{{\bf p}}{E} + N {\mbox{\boldmath$ \hat{\sigma}$}} +R \frac{{\bf p} \times \mbox{\boldmath$ \hat{\sigma}$}}{E} \right), 
\end{align}
where ${\bf P} =  {\langle{\bf J}\rangle}/{J}$ represents the average neutron beam polarization.
It has been pointed out in \cite{Sever06} that the Fierz interference term ($b\,m /\!E$) affects 
most of the correlation measurements using neutron spin asymmetry to extract correlation 
coefficients in a way, that the measured quantity becomes 
\begin{equation}\label{XFierz}
\tilde{X}=X/(1+\langle b\,m/E\rangle)
\end{equation}
 with $X=a,A,B,D \ldots$ and the averaging is performed over the observed $\beta$-spectrum.

\subsection{\label{S2_1}  Final state interaction and exotic couplings}

Following \cite{Lee56} the general Lorenz invariant interaction Hamiltonian density of nuclear 
beta decay can be written as:
 \begin{align}\label{Ham}
\nonumber
H =&(\bar{p}n)\big(\bar{e}(C_{S}\!+\!C'_{S}\gamma_{5})\nu\big)+
 (\bar{p}\gamma_{\mu}n)\big(\bar{e}\gamma_{\mu}(C_{V}\!+\!C'_{V}\gamma_{5})\nu\big)\\
\nonumber
&+{1}/{2} \,(\bar{p}\sigma_{\lambda \mu}n)
                            \big(\bar{e}\sigma_{\lambda \mu}(C_{T}\!+\!C'_{T}\gamma_{5})\nu\big)\\
\nonumber
&-(\bar{p}\gamma_{\mu}\gamma_{5}n)
                          \big(\bar{e}\gamma_{\mu}\gamma_{5}(C_{A}\!+\!C'_{A}\gamma_{5})\nu\big)\\
&+(\bar{p}\gamma_{5}n) \big(\bar{e}\gamma_{5}(C_{P}\!+\!C'_{P}\gamma_{5})\nu\big)+ H.c. ,
 \end{align}
where $C_{i}$ and $C'_{i}$ represent 10, in general complex coupling constants, which determine 
the symmetry properties of the weak interaction. 
In the minimal formulation of the SM only vector  and axial-vector interactions are present 
($C_{V}\!=\!C'_{V}\!=\!1$, $C_{A}\!=\!C'_{A}\!=\!\lambda$) and all other 
couplings vanish ($C_{i}\!= \!C'_{i} \!= \!0$; $i=S,T,P$). 
With these assumptions, both $R$ and $N$ vanish at the lowest order in neutron decay, but 
 acquire finite values when final state interactions are included:
 \begin{eqnarray}\label{NR_N}
N_{\text {FSI}} &\approx& -\frac{m}{E}\cdot\! A,\\
R_{\text {FSI}} &\approx& -\alpha_{\text {fs}}\frac{m}{p}\cdot\! A,
\label{NR_R}
 \end{eqnarray}
where $\alpha_{\text {fs}}$ is the fine structure constant.
For the energy distribution observed in the present experiment one obtains 
$N_{\text {FSI}}\approx0.068$ and $R_{\text {FSI}}\approx0.0006$. 
This means that the $R_{\text {FSI}}$ is far below 
the sensitivity of this experiment, while the finite value of $N$ should easily by measured.  

Allowing for a small admixture of  exotic couplings and keeping only terms linear in these 
couplings, one finds \cite{Jack57} that:
 \begin{eqnarray}\label{NR_N1}
N - N_{\text {FSI}} &\approx& \!-0.218\cdot\! \Re(S) + 0.335\cdot\! \Re(T), \\
R - R_{\text {FSI}} &\approx& \!-0.218\cdot\! \Im(S) + 0.335\cdot\! \Im(T),
\label{NR_R1}
 \end{eqnarray}
where 
\begin{eqnarray}\label{DefS}
S&= (C_{S}+C'_{S})/C_{V}, \\
\label{DefT}
T&= (C_{T}+C'_{T})/C_{A},
\end{eqnarray}
are the relative strengths of scalar and tensor interactions with respect to the dominant vector 
and axial-vector couplings, respectively.
Within these assumptions the coefficient $b$ can be expressed as \cite{Jack57}:
 \begin{eqnarray}
b &\approx& \!0.170\cdot\! \Re(S) + 0.830\cdot\! \Re(T),
\label{FirzTerm}
 \end{eqnarray}
and would affect the measured correlations following Eq.~(\ref{XFierz}).
However, the additional terms are of second order in the contributions of exotic couplings and 
can thus be neglected.

A non-zero value of the $R$ correlation in neutron decay would signal the 
existence of a non-vanishing contribution from imaginary couplings in the weak interaction, a new 
source of the TRV and, as a consequence, physics beyond the SM. 

The $N$  correlation depends on the real part of the same linear combination of scalar and 
tensor couplings as $R$.  
However, the discovery potential of its measurement is strongly suppressed by a significant 
contribution of uncertainties connected with the evaluation of the final state interaction. 

There exist very few measurements of $N$ and $R$ correlations in general 
\cite{Dan05,Abe04}, and only two in nuclear beta decays \cite{Sch83,Hub03}.

\section{\label{S3}  Experiment}

The key feature of the nTRV experiment is the ability to measure energies and to track 
over relatively long distances electrons from neutron decay.
This  allowed for efficient use of one of the world strongest polarized, cold neutron beams 
as a source of electrons from neutron decay and for application of efficient electron 
polarimetry based on Mott-backscattering \cite{Ban06}.
An additional advantage of this principle is the unique signature of relatively rare Mott 
scattering events which made them easily distinguishable off- but also on-line from an 
overwhelming background of electrons accompanying this, very strong, neutron beam.

Though a similar concept of electron detection has already been applied in the 
measurement of the neutron life time and the $A$ correlation coefficient 
\cite{Kos89,Lia97}, the present experimental setup outperforms the former ones, providing 
much more accurate reconstruction of both the electron  trajectories and their energies.

The experiment was performed  at the SINQ facility of the Paul Scherrer Institute, Villigen,
 Switzerland. 


\subsection{\label{S3_1}  Cold neutron beam}

\begin{figure}
\includegraphics[scale=0.34]{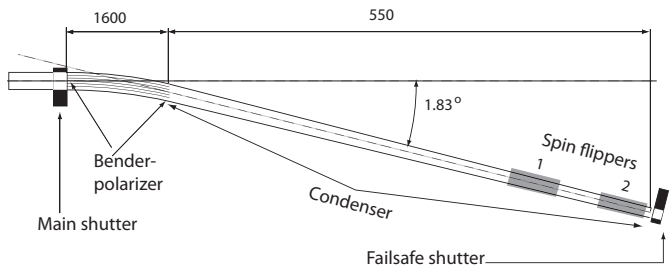}
\caption{\label{BeamLine} Schematic top view (not to scale) of the polarized cold neutron 
beam line arrangement. Cold neutrons from the liquid deuterium moderator enter from 
the left.} 
\vspace*{-1pt}
\end{figure}

A dedicated cold neutron beam line has been constructed for the present 
experiment at channel 51 of SINQ, leading directly to the cold moderator 
container.
The container was filled with about 20 liters of liquid deuterium at 25~K. 
Cold neutrons from the moderator were polarized in a 1.6 m long multichannel bender-polarizer 
\cite{Sche94}, and subsequently transported to the experimental area via a 
rectangular channel, farther referred to as condenser (Fig.~\ref{BeamLine}). 
Its  convergent, vertical walls matched the 80~$\times$~150~mm$^2$ entrance 
beam cross section with the about 5.5 m distant 40~$\times$~150~mm$^2$ exit.
The condenser's main role was to increase the neutron density at the experiment, 
and to separate this area from a large background of fast neutrons and gammas produced in 
the SINQ interior and in the polarizer. 
To minimize background and neutron losses due to interactions with gas,  
the polarizer and condenser were enclosed in a vacuum chamber with  180~$\mu$m and~125 $\mu$m 
thick zirconium entrance and exit windows, respectively.  

The application of carefully chosen, different kinds of supermirrors 
\cite{Boni97, Sche96, Sche99}, with a critical reflection angle up to 3.3 times larger than 
that of natural nickel, applied in the polarizer and covering the walls of the beam line, 
allowed for maximum neutron polarization and transmission efficiency.
A vertical spin guiding magnetic field was maintained along the beam line by a combination 
of permanent magnets and iron plates. 
The magnitude of this field was approximately constant along the polarizer and over a large 
part of the condenser, decreasing only at the spin-flippers position to create field 
gradients, necessary for spin-flipper operation.
Two adiabatic radio frequency spin-flippers, mounted around the last section 
of the condenser, were used to reverse the orientation of the neutron beam 
polarization at regular time intervals, typically every 16~s.

The experimental area was  shielded from the background produced 
in the neutron guide and polarizer by a 0.5 m thick concrete wall with 
inserted boron-lead collimator (Fig.\ \ref{setup}).
The 1.3 m long, multi-slit collimator defined the beam cross section to
40$\times$150 mm$^{2}$ at the entrance of the Mott polarimeter. 
In order to minimize neutron scattering and capture, the entire beam line,
 from the collimator to the beam dump, 
was enclosed in a chamber lined with $^{6}$LiF polymer and filled with pure 
helium at atmospheric pressure. 

The total flux of the collimated beam was typically about 10$^{10}$ neutrons/s. 
The beam divergence was 0.8$^{\circ}$ in the horizontal and 1.5$^{\circ}$ in the 
vertical direction. 
A detailed study of the beam polarization at the position of the experimental setup was 
performed with a polarization analyzer based on the bent supermirrors concept, analogous to 
the one used in the polarizer. 
The obtained results revealed a maximum polarization of 95\% in the beam center and its 
strong dependence on the position and on the inclination angle with respect to the beam axis.
This feature hindered a reliable evaluation of the average polarization integrated over the 
whole beam fiducial volume. 
The adopted solution was to measure, in parallel to the main correlation experiment, and with the 
same beam and detector, also the neutron $\beta$-decay asymmetry.
As the asymmetry parameter  is known with a high precision from other experiments, 
this approach allows the extraction of the average beam polarization, 
while automatically accounting for the complicated beam phase space and the detector acceptance.  
The obtained results are listed in Table \ref{tab:table2}.

A more detailed description of the design, operation and performance of the cold neutron 
beam line can be found in Ref.~\cite{Zej04}.

\subsection{\label{S3_2}  Detector setup and performance}

\begin{figure}
\includegraphics[scale=0.75]{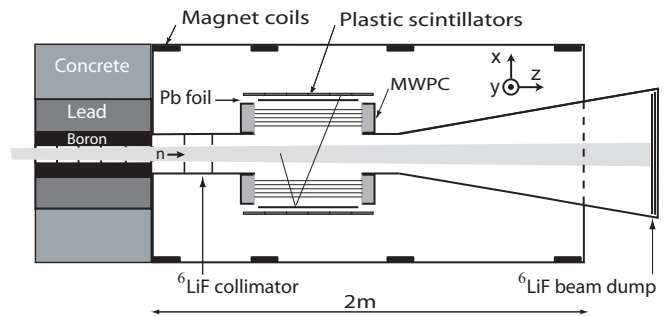}
\caption{\label{setup} Schematic top view of the experimental setup.  A~sample projection 
of an electron V-track event is shown. The coordinate system used throughout this 
paper is indicated.}
\vspace*{-7pt}
\end{figure}

The Mott polarimeter consists of two identical modules, 
arranged symmetrically on both sides of the neutron beam (Fig.\,\ref{setup}). 
The whole structure was mounted inside a large volume dipole magnet 
providing a homogeneous vertical holding field of 0.5 mT within the beam
fiducial volume. 
Going outwards from the beam, each module consists of a multi-wire proportional 
chamber (MWPC) for electron tracking, a removable Mott scatterer (Pb foil)
 and a plastic scintillator hodoscope for electron energy measurement.

The main requirements which shaped the design of the MWPC were the minimal 
energy loss and multiple scattering of low energy electrons and the possibly small 
cross section for conversion of gamma quanta into electrons, which would pose a 
dangerous background source. 
To fulfil these conditions a unique combination of special features was implemented:
\begin{itemize}
\item
The readout of anodes and cathodes allowed reducing the total thickness of the MWPC by a factor 
of two. 
\item
Very light gas mixture based on helium, isobutane and methylal (90/5/5), which 
nevertheless assured stable working condition at anode voltages of about 1800 kV.
\item
Thin, 25 $\mu$m Ni/Cr (80/20) wires at 5 mm and 2.5 mm pitch for anodes and cathodes,
respectively,
\item 
Very thin entrance and exit windows made of 2.5 $\mu$m aluminized Mylar foil.
\end{itemize}
Each chamber contained five planes of anodes (horizontal wires) and five planes of measuring 
cathodes (vertical wires), with active areas of 50$\times$50~cm$^{2}$.
The distance between anodes and cathodes was 4~mm and between the consecutive anode planes 
16~mm.
The average efficiency of a single plane was about 98\%  and 97\% for anodes and
 cathodes, respectively.

The time measurement of individual wire hits with respect to the reference signal from the 
scintillator depends on the spatial density distribution of the primary ionization. 
Unlike in drift chambers, with electrically separated drift cells and a sense wire in 
their center, the adopted MWPC geometry does not allow the use of the drift time of primary 
ionization electrons to improve the position resolution of a single plane. 
However, the time information was used to improve the reconstruction of the cluster centroid for 
the cases in which more than two neighboring wires have responded.   
It was also used to check whether large clusters are not formed by two overlapping 
smaller clusters. 
This allowed improving the position resolution and the double track resolution of 
a single plane, and was of special importance for cathodes, with an average cluster 
size of about 1.9~hits.
The position resolution of a single plane, obtained from the distribution of reconstruction 
residua, was about 1.2~mm and 1.7~mm r.m.s. for anodes and cathodes, respectively.
Better double track and position resolution of anodes lead to more precise and more 
efficient reconstruction of trajectories in the vertical coordinate, relevant for the $R$ 
correlation coefficient measurement. 

The  scintillator hodoscopes were optimized for the detection of electrons with energies up 
to about 1.8~MeV. 
This allowed the distinction of electrons originating from neutron decay from more energetic 
background electrons and played a crucial role in the background subtraction procedure.
Each hodoscope consisted of six 1 cm thick, 10 cm wide and 63 cm long plastic 
scintillator slabs.
Two XP3330 photomultipliers were coupled optically to both ends of the scintillator via 
short light guides of optimized shape.
This solution allowed for reconstruction of the total electron energy with 33~keV resolution 
at 500~keV and 48~keV at 976~keV (see Fig.~\ref{Bi207}). 
The asymmetry of the light signal collected at both ends of the scintillator slab 
allowed the determination of the vertical ($y$) hit position 
with a resolution of about 6 cm, while the segmentation of the hodoscope in the 
horizontal direction provided a crude estimate of the $z$-coordinate.
Matching the information from the precise track reconstruction in the MWPC 
with that from the scintillator hodoscope considerably reduced the background and 
random coincidences.

Fast pulses from the hodoscope were also used in a trigger logic and provided the time reference 
signal for the MWPC wire readout.

\subsection{\label{S3_3}  Mott scatterer}

\begin{figure}[htb]
\includegraphics[scale=.28]{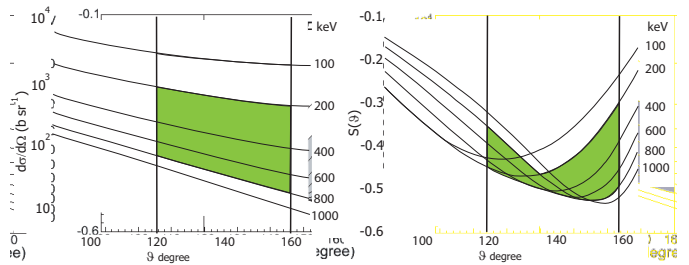}
\caption{\label{Mott}Angular distributions of the Mott scattering cross section (left panel) 
and the Sherman function (right panel) for electrons on natural lead. Shaded areas show 
parameter regions relevant to the present experiment.}
\end{figure}

Scattering in the field of a spin-less nucleus of electrons polarized in the direction 
perpendicular to the scattering plane reveals a left-right asymmetry due to the spin-orbit 
term present in the interaction potential.
The purely electromagnetic nature of this process (effects of the weak neutral currents can 
be safely neglected in this case) guaranties the exclusive sensitivity to the transverse 
polarization of the incoming electron.
The resulting asymmetry is proportional to a product of the transverse polarization component 
of the incident electron beam and the target-specific analyzing power of the scatterer.
The experiment presented here exploits particularly favorable conditions existing for 
electron scattering on high $Z$ nuclei (lead) at large backward angle, where the 
Sherman function (analyzing power of a single nucleus \cite{Sher56}) reaches its highest 
value, and the Mott scattering cross section is still appreciable (Fig.~\ref{Mott}).

\begin{figure}[htb]
\includegraphics[scale=.43]{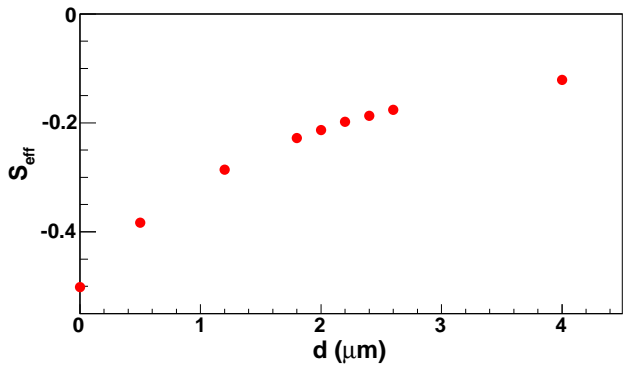}
\caption{\label{S_ef_thick} Simulated effective analyzing power of lead for 400~keV electrons 
backscattered at an angle of 140$^{o}$ as a function of the lead layer thickness. Electrons 
incident perpendicularly to the foil.}
\end{figure}

For a real scatterer one has to take into account the inelastic multiple scattering with 
atomic electrons and, although being much less likely, plural Mott scattering at moderate 
angles which eventually can mimic a single Mott-backscattering event.   
These effects can significantly deteriorate the initial analyzing power and affect the data 
by their substantial dependence on the thickness of the scatterer (c.f. Fig.~\ref{S_ef_thick}) 
and on the incidence angle with respect to the foil surface.

In order to obtain the effective analyzing power of the 
scatterer, Monte-Carlo simulations were performed using the Geant 4 simulation 
framework \cite{GEANT4} and following guidelines presented in Refs.~\cite{Sal05,Kha01}. 
This approach takes advantage from accurate theoretical calculations of the Sherman 
functions, properly accounting for atomic structure and nuclear size effects, as well as for 
effects due to electrons interaction with the medium.
The accuracy of these calculations has been estimated to be better than 2\% and was verified 
by comparison with two experimental data sets: at 
low (120 keV \cite{Ger91}) and high (14 MeV \cite{Srom99}) electron energies.

In the early phase of this experiment (2003 - 2004) a 1~$\mu$m thick lead layer evaporated on 
a 2.5 $\mu$m thick Mylar foil was used as the Mott-scatterer. 
The bulk of the data, however, was collected in 2006 and 2007 with about twice as large 
surface density of the lead scatterer.
Even this foil was almost transparent to the incident electrons from neutron decay, so that 
more than 99\% of them penetrated through the foil without sizable interaction with 
lead nuclei.

Due to the Mott-foil manufacturing process, the thickness of the lead layer was not 
perfectly uniform. 
Also the illumination of the foil by electrons at the experiment position was not uniform.
This was included in the systematic uncertainty in the analysis presented in Ref.~\cite{Koz09}. 
In order to decrease this uncertainty and to enhance the reliability of the obtained 
results, precise scans of the lead surface density distribution of both 2 $\mu$m Mott 
scatterers were performed using photon-induced characteristic radiation \cite{Koz10}. 

\begin{figure}[htb]
\includegraphics[scale=.43]{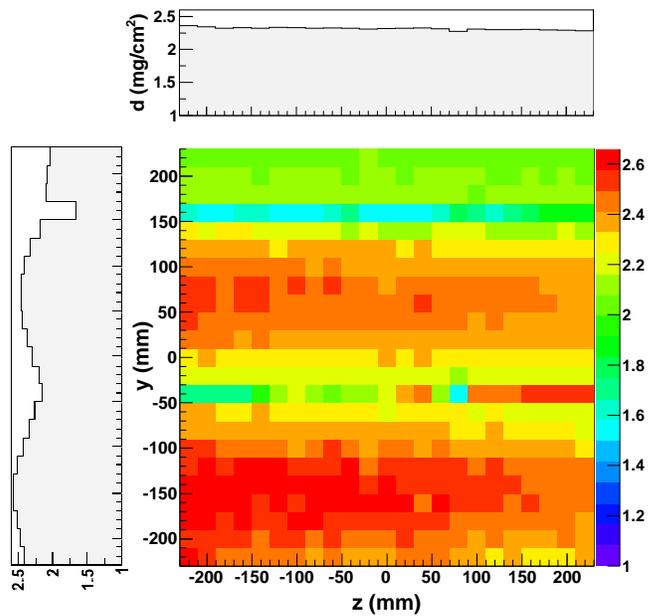}
\caption{\label{DensMap} (Color online) Surface density map of one Mott-target with its 
projection on both $y$ and $z$ axes. The deep minimum around  $y$=150~mm is due to imperfect 
matching of individual sheets used for the 
Mott-target fabrication. Smooth changes which are visible in $y$-projection are 
the relics of the evaporation process used for manufacturing of a single sheet.}
\end{figure}

The resulting maps of the lead layer surface density measured with an absolute accuracy of 
about 55 $\mu$g/cm$^{2}$ (Fig.~\ref{DensMap}), together with the Monte-Carlo simulated 
multidimensional effective analyzing power data 
and the distributions of the 
reconstructed electron vertices on the scatterer were used to obtain the final average 
effective analyzing power values.

The new values are by about 7\% smaller than those used in Ref.~\cite{Koz09}, with relative 
uncertainty reduced from 9\% to about 3\%.

\subsection{\label{S3_4}  Detector electronics and hardware trigger}

The entire electronics coupled directly to the detectors (sense wires of MWPC and 
photomultipliers) responsible for signal amplification, discrimination and derivation 
of the most important early stage of the hardware trigger, was designed and built 
specially for this experiment.  
The details of the implementation are described in Ref.~\cite{Hil01}, here only the main 
concept of the trigger is presented.

\begin{figure}[htb]
\includegraphics[scale=.35]{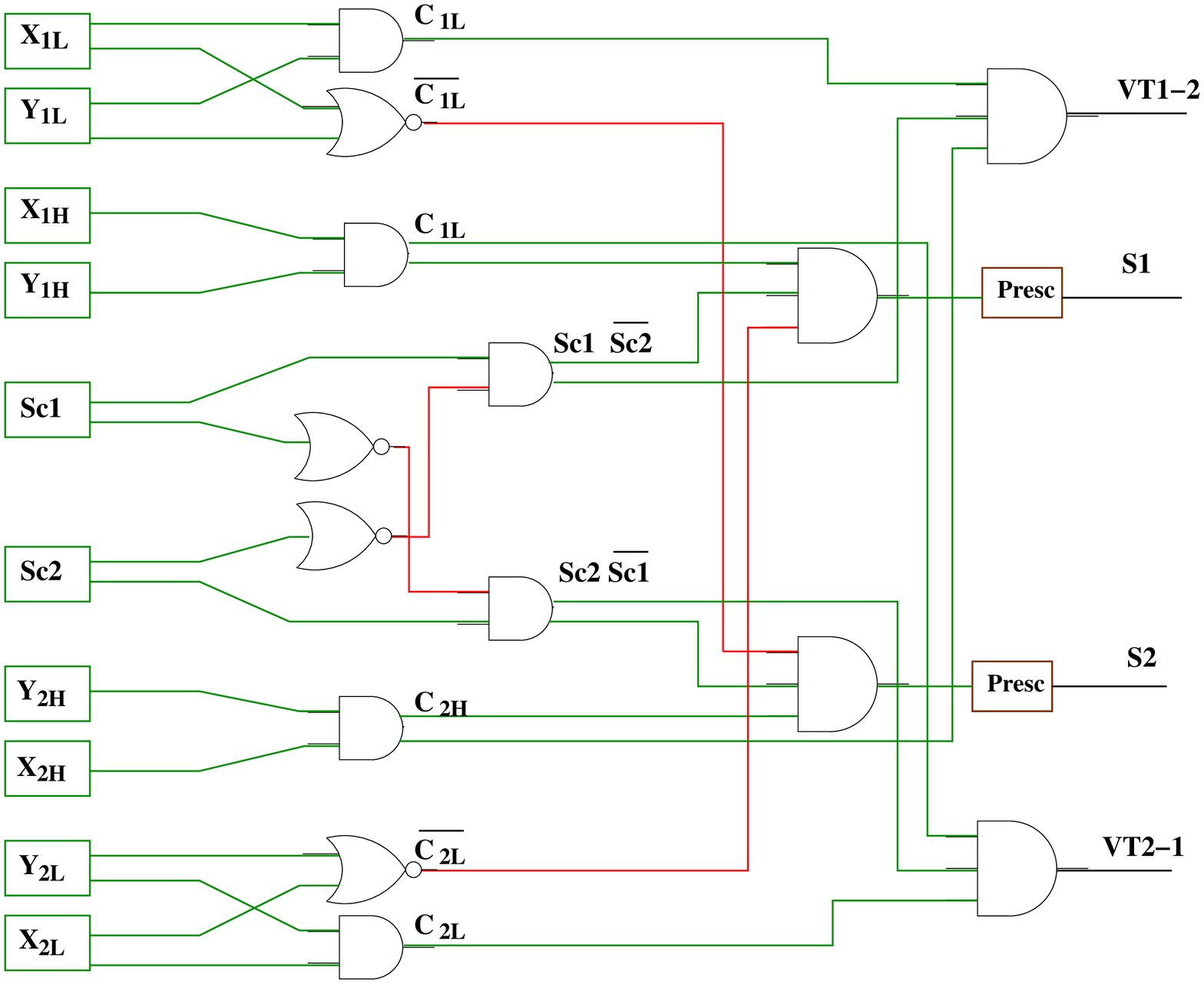}
\caption{\label{HardTrig}(Color online) Simplified diagram of the hardware trigger. 
Shown is the logic of V-track (VT1-2, VT2-1) and single track events (S1, S2) constructed 
from scintillator signals (Sc1/2) and chamber multiplicities low and high ($X_{1/2\,L/H}, 
Y_{1/2\,L/H}$). }
\end{figure}

The hardware trigger was built such as to collect virtually all events 
belonging to each of two classes:
\begin{itemize} 
\item
VT1-2 and VT2-1 in Fig. \ref{HardTrig} -- Mott scattered electrons with two track segments 
on one side and one segment accompanied by a scintillator hit on the opposite side, 
 further referred to as ``V-track'' (an example is shown in Fig.~\ref{setup}), used for the 
determination of the electron transverse polarization,
\item
S1 and S2 in Fig.~\ref{HardTrig} --- ``single-track'' events with only one reconstructed 
track segment on the hit scintillator side, used for precise evaluation of the average beam 
polarization.
\end{itemize} 

In order to enhance the selectivity of the trigger, two plane multiplicity signals have been 
constructed separately for anodes ($Y$)  and cathodes ($X$) of each detector side.
High plane multiplicity ($X_{iH}$, $Y_{iH}$, where $i$ indicates the side of the 
detector), was relevant for the detector side which reconstructed two track segments in the 
MWPC, and low multiplicity ($X_{iL}$, $Y_{iL}$) was required for the detector side with only 
one segment.
Taking advantage of the small drift cell size and using fast OR circuits implemented on the 
discriminator boards of each MWPC plane, both signals were generated as early as 80 ns 
after the fastest scintillator pulse.
This concept allowed the data acquisition to collect V-track events without losses, 
admitting only the most promising candidates for single track events, with 
the possibility of their further reduction by the prescaler module (Fig.~\ref{HardTrig}).

Typically the low (high) plane multiplicity signal was set when more than two (three) 
planes of the same electrode type registered at least one hit wire.
With this setting the observed rates were about 800 and 8000 Hz for V-track and for single 
track events, respectively.
In order to keep the dead time at the acceptable level below 10\%, the single track 
event rate was prescaled by a factor of two.

\subsection{\label{S3_4}  Data acquisition}

For each trigger the information including pulse heights and time measurements with respect 
to the fastest hodoscope signal for all hodoscope hits, and time measurements for all MWPC 
wire hits were digitized in FERA compatible ADC (LeCroy 4300B) and TDC (LeCroy 3377) modules.
In order to enhance the data throughput of the standard FERA bus, a custom CAMAC  
module has been applied (the FERA Tagger/Extender). 
It allowed the separation of  readout electronics into two logical FERA subsystems with separate 
gates \cite{Kis10}. 
The data from each subsystem were transferred in parallel to two pairs of 
VME hosted memory modules working in flip-flop mode. 
The memories were read out via a VME data-bus controlled by a RIO2 processor running MBS 
data acquisition software \cite{Kur99} installed on the real-time operating system LynxOS. 
An important role of the data acquisition program was the generation of periodic interrupts, 
typically every second, used for read-out of monitoring scalers and for setting the spin 
flippers controlling the beam polarization. 
This software was also responsible for final logging of the data on the external mass storage 
and for sending a fraction of the data to the back-end computer for monitoring purposes. 

The  average data flow rate was about 1~MB/s, which amounted to about 15\% of the maximum 
achieved data throughput of the system.

\section{\label{S4}  Data Analysis}

The final result of this experiment is dominated by the last and the longest data 
collection period.
The numbers quoted in this chapter apply to this period,  however, all available data were 
consistently analyzed in an analogous way, and the obtained results are presented and 
used in the calculation of the final average.

\subsection{\label{S4_1}  Data reduction}

One of the most important features of the data analysis is its hierarchical structure.
The raw data collected during the experiments, coded in a compact format specific to the 
individual electronic modules, were first converted to ``physical'' format (trajectory 
segments, deposited energy in hit scintillators) with subsequent verification of the 
on-line trigger conditions. 
All parameters specifying these conditions were set in such a way that ``good'' events must 
not be removed regardless of their origin ({\em i.e.} from- or off-the-beam). 
This allowed the preselection of interesting event classes thereby reducing the amount of data 
to be processed at the next stage of the analysis by a factor of about twelve in the case of 
back-scattered events, and by about 50\% in the case of single track events.
 
The second step prepared the final selection of events using tighter conditions, thus further 
reducing the amount of data but still allowing for some freedom in setting the most crucial 
parameters (listed in Section \ref{SystErr}).

\subsection{\label{S4_2}  Calibration of scintillator hodoscopes}

The energy and position calibration of the hodoscopes was performed typically once a week 
using conversion electrons from a $^{207}$Bi source.
A movable support driving the source in $y$ and $z$ directions within the  symmetry plane 
of the detector ($x=0$) between the chambers was used to provide uniform illumination of 
the entire detector. 

\begin{center}
\begin{figure}[htb]
\includegraphics[scale=.44]{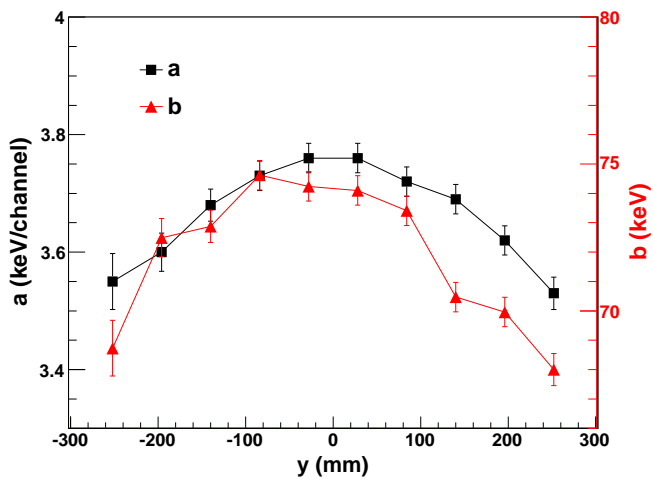}
\caption{\label{EnCal}(Color online) Energy calibration for one of the scintillators (see 
Eq.~(\ref{EnCalEq})). The residual position dependence of calibration coefficients $a$ and $b$ 
are due to non-uniform light attenuation along the scintillator bar.}
\end{figure} 
\end{center}

The reconstructed electron trajectories allowed the identification of the hit position 
along the individual scintillator and the correction for path-length dependent electron 
energy losses.
As a consequence, it was possible to calibrate separately relatively short sectors of a 
scintillator and to obtain energy calibration specific for the position ($y$) at which 
the energy was deposited in the scintillator.
The reconstructed deposited energy $E$ was assumed to be a linear function of the 
scintillator response function ($g$):
\begin{equation}\label{EnCalEq}
E = a_{i}(y) \, \, g(E_{d}, y) + b_{i}(y), \quad i=1\ldots 12.
\end{equation}
The function $g$  was defined as the geometrical mean of the pulse heights 
recorded by the ``up'' and ``down''  photomultipliers ($c_{u},\,c_{d}$):
\begin{equation}
g = \sqrt{c_{u}\,c_{d}}.
\end{equation}
With this definition, and assuming uniform light attenuation along 
the scintillator, the $g$-function should be proportional to the deposited energy ($E_{d}$) and 
should not depend on $y$.
Small deviations from this assumption were compensated by the position sensitivity of the 
 calibration coefficients $a_{i}$ and $b_{i}$ (Fig.~\ref{EnCal}).

\begin{center}
\begin{figure}[htb]
\includegraphics[scale=.46]{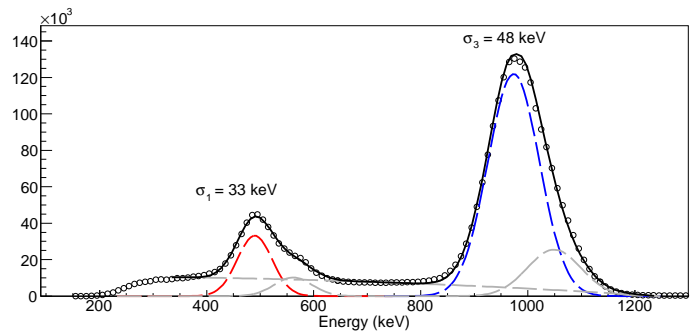}
\caption{\label{Bi207}(Color online) Measured  $^{207}$Bi electron energy spectrum (circles) 
together with its decomposition into the four most important electron conversion lines (482, 558, 
976 and 1052~keV) and a smooth background. The fitted widths correspond to the energy resolution 
of the scintillator.}
\end{figure}
\end{center}

The energy resolution can be deduced from the reconstructed  $^{207}$Bi electron energy 
spectrum (Fig. \ref{Bi207}).
It roughly follows the primary photon statistics and amounts to 33 keV around 500 keV. 
The associated systematic uncertainty has been estimated to 5 keV. 
In order to decrease the energy spread caused by different energy losses on the way 
between the electron creation point (assumed to be in the symmetry plane of the detector) and 
the scintillator hodoscope, only events with similar path-length have been selected for the 
calibration.
 
\begin{center}
\begin{figure}[htb]
\includegraphics[scale=.40]{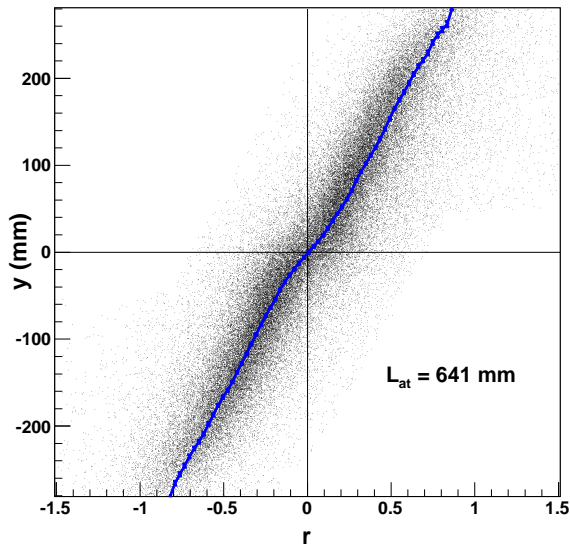}
\caption{\label{PCal0}(Color online) A sample position calibration $\mathcal{Y}(r)$ of 
one scintillator bar (solid line) superimposed on the experimental events distribution. 
An event is represented by the $r$ asymmetry (Eq.~\ref{LogAsy}) and hit position along the 
scintillator reconstructed from the MWPC information.
The non-linearity due to nonuniform light attenuation along the scintillator is clearly visible. 
$L_{at}$ corresponds to the average light attenuation length, Eq.~(\ref{LogAsy}).}
\end{figure}
\end{center}

Using once again the uniform light attenuation assumption one can show that the signal asymmetry 
defined as: 
\begin{equation}\label{LogAsy}
r=\ln{(c_{u}/c_{d})}= \frac{2\,y}{L_{at}}
\end{equation}
should be directly proportional to $y$ and should not depend on $E_{d}$.
Figure \ref{PCal0} presents the $y$-component of hit positions on the hodoscope reconstructed 
from the MWPC information and plotted against the $r$ asymmetry. 
The correlation between both observables is obvious.
The large width of the band reflects the modest position resolution of the scintillator. 
The average $r$ at a given position $y$ was used to obtain the position calibration of an 
individual scintillator  $\mathcal{Y}(r)$.
It should be noted that this calibration was not perfectly linear, what again indicates a small  
departure from the uniform light attenuation assumption.

\subsection{\label{S4_3}  Data selection  and event reconstruction}

The first step of the data selection procedure is a fine tuning of the coincidence time windows 
in which hodoscope and MWPC hits are accepted in order to reduce event contamination with 
accidental coincidences.
The resulting width of the time window used for the MWPC hits (180 ns) accounts for the maximum possible 
drift time of primary ionization electrons in the drift cell and for the time walk due to 
variations of the signal rise time.
As the time information from scintillators was affected only by the time-walk, the length of 
the corresponding gate for the hodoscope signals could have been much shorter and amounted to 
50~ns. 

For a valid hodoscope hit the coincidence between the photomultipliers attached to both ends of 
one scintillator was required. 
In general only one such hit was allowed in both hodoscopes. 
Exception was made for the cases when two neighboring scintillator slabs responded. 
This allowed the selection of electrons reaching the hodoscope at the scintillators edges 
which deposited their energy partially in two neighboring scintillators.
Those events played an important role in the determination of the position resolution 
of track reconstruction at the hodoscope position, based on the position information from 
the anodes. 

As already mentioned in section \ref{S3_2}, the information from the MWPC was reduced to the 
cluster centroids. 
The average cluster size was about 1.4 and 1.9 wires for anodes and cathodes, respectively.
Clusters consisting of more than 4 consecutive wires were investigated in order to check whether 
they were formed by two overlapping clusters.
The splitting condition was derived from the analysis of the time information of all hit 
wires belonging to this cluster and relied on the presence of two significant minima in 
the corresponding time distribution. 
As double clusters are naturally present in vertex topology (with their relative distance 
decreasing for wire planes closer to the Mott target), this filter was particularly 
important to increase the detection efficiency of V-track events. 
Clusters larger than 16 wires have been rejected as a possible electronics noise (16 channels 
were grouped in one preamplifier-discriminator card). 

Straight lines were then fitted to the obtained hit patterns separately in anodes 
and in cathodes using a combinatoric algorithm and the minimum $\chi^{2}$ criterion.
To be accepted a track projection had to be detected in at least three wire planes. 
An event was rejected if at least one complete track segment (seen in both projections) 
was registered in excess of the expected number of segments, that is exactly one segment 
at the hit hodoscope side and, in the case of V-track 
events, additionally two segments at the opposite detector side.

In order to reduce background consisting of electrons produced in the solid parts of the 
detector, only track segments whose prolongations were contained within the active area 
of the opposite MWPC were accepted.
This allowed using the two innermost planes of the opposite chamber as a veto detector and, 
as a consequence, to confine significantly the volume of possible electron origins.

The reconstructed track segments were confronted with a set of conditions checking the 
consistency of the reconstructed event.  
The extension of the segment reconstructed in the MWPC at the hit hodoscope side should point, 
within a given tolerance, to the hit scintillator slab and should match the $y$-position 
reconstructed from the ''up-down'' asymmetry of the corresponding pulse height signals.
In the case of V-track events  two lines in each projection were expected at the detector 
side which registered the Mott scattering vertex.
The $x$-coordinate of the vertices reconstructed in both projections should match within a 
tolerance given by the MWPC angular resolution. 
Similarly, matching in both projections was required between the segment reconstructed at 
the hit scintillator side and one of the two segments at the opposite side.
No scintillator hit in the vicinity of the reconstructed vertex was allowed for a valid Mott 
backscattering event.

Additional event classes, absent in the previous publication (Ref.~\cite{Koz09}) and included 
in the analysis presented in this paper, consist of cases for which the Mott scattering vertex 
was detected only in one projection ($y$-$z$ or $x$-$z$). 
Particularly interesting are events scattered close to or in the vertical plane. 
For these the projections of both segments in the $x$-$z$ plane overlap and are 
reconstructed as only one line.
The signature of such event is weaker: since one vertex is missing one important 
matching condition (vertices $x$-coordinates) drops out. 
Some compensation of this relaxation was achieved by increasing from three to four the 
threshold for the minimum plane multiplicity of the accepted line fitted in this projection. 
This is justified since each wire along such double track collects twice as much charge as in 
the normal case.
It should be noted that the $x$-$z$ projection was measured by cathodes i.e. electrodes with 
significantly worse double track resolution than anodes.
This increased the number of such events, further referred to as ``vertical single vertex 
events''.
This  event class exhibits maximal and exclusive sensitivity to the $R$ correlation, on the 
average two times higher than the primary double-vertex event class (see discussion of 
geometrical form-factors in Sect.~\ref{S4_7}). 

The analogous event class, with one vertex reconstructed by cathode planes accompanied 
by only one track segment reconstructed in anodes, is much less numerous than the vertical 
single-vertex event class. 
This is due the lower efficiency, position resolution and double track resolution 
of cathodes. 
Those events are sensitive almost exclusively to the $N$ correlation coefficient.

In the last step of the event reconstruction process each event was assigned to one of several 
event classes determining its role in the further analysis.
The most important categories were related to:
\begin{itemize}
\item
external conditions: state of beam polarization and presence of the Mott target,
\item
event geometry: from- and off- the beam, from- and off- the Mott-foil. 
\end{itemize}

The total numbers of reconstructed events are listed in Table \ref{tab:Stat}, separately 
for all data collection periods.
\begin{table}[htb]
\caption{\label{tab:Stat}Average lead surface density $d$ of the used Mott target and total 
numbers of reconstructed single track events $S$, double vertex events  $VV$ and events with one 
vertex in anodes $Va$ and cathodes $Vc$ for all data collection periods.}
\vspace{2mm}
\begin{ruledtabular}
\begin{tabular}{cccccc}
Run  &$d$ ($\mu$g/cm$^{2}$)&$S\!\times10^{3}$\!&$VV\!\times\!10^{3}$&$Va\!\times\!10^{3}$&$Vc\!\times\!10^{3}$ \\
\hline
2003 & 1.13 $\pm$0.16 & 12000 & 19     &&   \\
2004 & 1.13 $\pm$0.16 & 43000 & 74     &&   \\
2006 & 2.46 $\pm$0.05 & 28000 & 312    & 106& 28 \\
2007 & 2.46 $\pm$0.05 & 334000& 1750   & 711& 248 \\
\hline
Total &&417000&2152&817&276\\
\end{tabular}
\end{ruledtabular}
\end{table}

\subsection{\label{S4_4}  Effects of magnetic field}

All matching conditions discussed above as well as the event geometry were affected by the 
magnetic field. 

In order to attain the required uniformity of the spin holding field in the beam fiducial 
volume, the entire detector was immersed in a constant, large volume magnetic field, produced by 
a magnet consisting of two soft iron plates and eight iron core coils (Fig.\ref{setup}).
Special care has been taken in order to 
shield all photomultipliers against the influence of this field.
This has been achieved with a double layer of mumetal shielding around each photomultiplier.

The effect of the magnetic field on the detected electrons is twofold. 
Their spins precess at the Larmor frequency and their trajectories are bent so that 
their $x$-$z$ projection becomes an arc.
For the measurement of the transverse polarization of electrons those effects are 
potentially dangerous, since the polarization is produced and analyzed at distant locations 
(neutron decay and Mott-scattering).
However, since the electron $g$-factor is almost equal to 2, the spin precession almost 
exactly follows the momentum rotation. 
The maximum remnant effect, due to the ``$g-2$'' factor, is well below one arc minute in this 
experiment and is therefore irrelevant for the achieved accuracy. 

There are, however, other consequences of the bending of the electron trajectories.
Part of them can be accounted for while others are discussed in order to demonstrate 
why they do not influence the final result. 

The deviation $\rho$ of the electron trajectory from the straight line depends strongly on 
the distance $d$ traveled by the electron in the magnetic field.
Neglecting energy losses it can be approximated as:
 \begin{equation}\label{Effect}
\rho \approx d \tan\left[\frac{1}{2} \arcsin\left( \frac{d}{R}\right)\right],
 \end{equation}
where $R$ is the curvature radius which, for the lowest electron energy detected in this 
experiment, amounts to about 3~m. 
From this it follows that in the worst case the deviation of the real trajectory from 
a straight line within one MWPC is below 0.5 mm.
Considering the average size of clusters in cathodes ($\approx$14~mm), such a small correction 
can be neglected and one may safely use a linear fit to the data.
This is not the case if one considers prolongations of the obtained lines to the Mott 
target, to the scintillator hodoscope or to the opposite MWPC. 
Then the effects can be substantial, but knowing the electron energy and the magnetic 
field strength they can be accounted for.

A strict correction would require taking into account continuous energy losses along the 
electron path.
However, in view of the much larger effect of electron multiple scattering, this correction 
has been simplified by using the average electron energy along the considered path segment. 
Subsequently the line fitted in the MWPC was treated as a tangent to the circular trajectory 
with a radius corresponding to this average energy. 
Trajectories obtained in this way were then used to calculate extrapolated electron 
positions and incidence angles at beam, hodoscope, Mott target, etc.

\begin{figure}[htb]
\includegraphics[scale=.32]{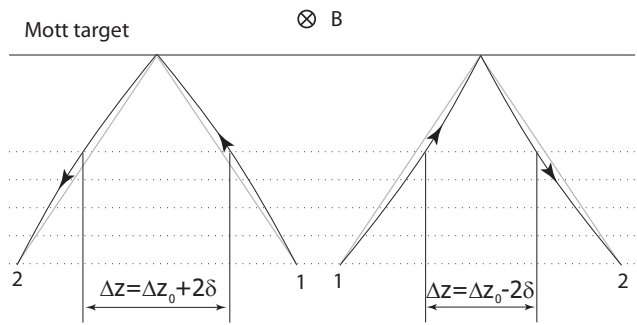}
\caption{\label{ConcaveConvex} Convex and concave vertices. $\delta$ corresponds to the 
displacement of the electron trajectory due to the magnetic field, measured along the 
$z$-axis at each wire plane.}
\vspace*{0pt}
\end{figure}

Surprisingly, even a weak magnetic field can have a significant influence  on the efficiency of 
V-track reconstruction.
In order to understand this effect two kinds of V-tracks must be introduced. 
In the following they will be referred to as convex and concave (Fig.~\ref{ConcaveConvex}).
For convex  V-tracks the distance between both trajectories measured within each wire 
plane (along the $z$ direction) is always larger than for its concave analog. 
This difference reaches its maximum around the outermost planes, closest to the Mott target, at 
the place where  the separation of clusters belonging to both arms of the vertex 
reaches its minimum. 
This causes a difference in the probability for two clusters in the same plane to overlap and, 
as a consequence, increases the detection efficiency of convex V-tracks and decreases this 
efficiency for concave V-tracks. 
Of course the same effect, however with decreasing significance, occurs also in other 
MWPC planes. 

Another effect may be described as focusing (for convex) or defocusing (for concave) of the 
long arm of a V-track on the active area of the opposite detector. 
It acts coherently with the previous one, further increasing (decreasing) the detection 
efficiency for convex (concave) V-tracks. 

In order to reduce the impact of all effects induced by the guiding magnetic field, 
its magnitude has been reduced from 1~mT, used between 2003 and 2006, to 0.45~mT in the 
2007 data taking period.
Nevertheless the effects persist and can readily be observed (see e.g. Figs.~\ref{FinFoutMF}, 
\ref{AlfaAng}).

\subsection{\label{S4_5}  Background correction }
              
Two kinds of background have been taken into account and corrected for. 
The first, further referred to as the ``off-beam'' background is present in single 
track and in V-track event classes, while the second, ``foil-out'' background, applies only to 
the V-track events.

\begin{figure}[htb]
\includegraphics[scale=.44]{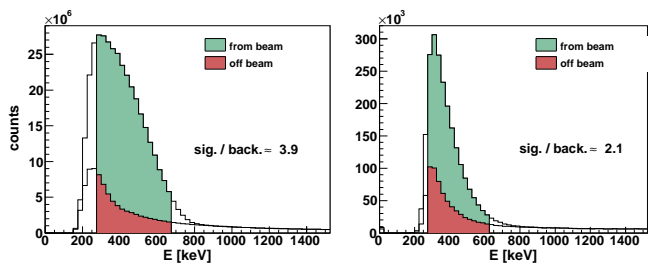}
\caption{\label{En_From_Out}(Color online) Energy distributions of signal (green) and 
background contributions (red) for single track events (left) and double vertex event class 
(right). Shaded areas indicate the ``signal'' energy range used in the final calculations.}
\vspace*{0pt}
\end{figure}

For the ``off-beam'' background the number of electrons not originating from the free 
neutron decay, 
was determined by comparing energy spectra of two event classes: (i) events for which the 
reconstructed electron trajectory crossed the neutron beam volume (``from beam'') and 
(ii) events for which the electron origin was outside the neutron beam (``off beam''). 
The procedure relies on the assumption that the spectral shape of the background is 
the same for both event classes, while the characteristic neutron $\beta$-decay spectrum with 
end-point energy of 782 keV is present only in the ``from beam'' class (Fig.~\ref{En_From_Out}).
This allows scaling the ``off beam'' background distribution such that it matches the high energy 
part of the  ``from beam''  energy spectrum. 
For this assumption to hold, the ``off beam'' range has to be carefully chosen for both: 
the inclination angle and the extrapolated origin of the tracks at the opposite detector side \cite{Ban06}. 
These conditions have to account for the angular resolution of the MWPC, the beam density 
distribution and its divergence.
The better signal to background ratio obtained for single track events can be attributed to 
their much larger number what allowed for a much tighter setting of all geometrical cuts.
 
The validity of this background subtraction method was verified by comparing background-corrected 
energy spectra with the simulated  $\beta$-decay spectra in which  energy losses and detector 
resolution were taken into account.

\begin{figure}[htb]
\vspace*{-7pt}
\includegraphics[scale=.44]{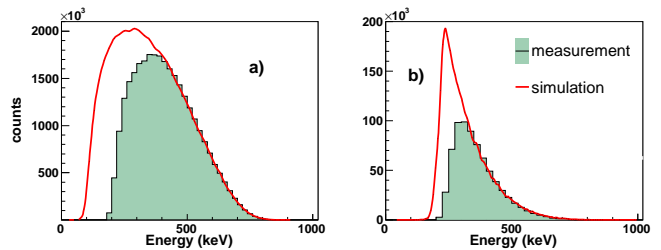}
\vspace*{-7pt}
\caption{\label{BetaSing}(Color online)  Background-corrected experimental energy 
	distributions (shaded areas) of (a) single-track and (b) double-vertex V-track 
        events 	compared with simulations.}
\vspace*{-5pt}
\end{figure}
\begin{figure}[htb]
\vspace*{-7pt}
\includegraphics[scale=.44]{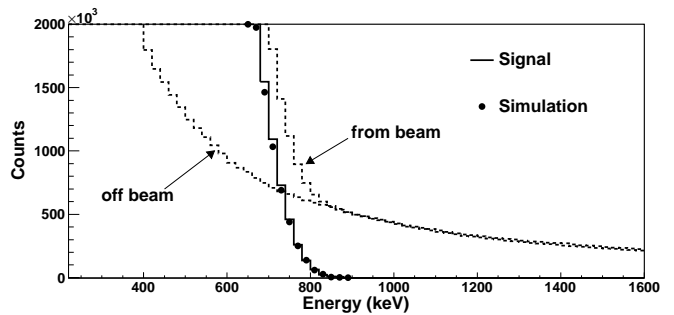}
\vspace*{-7pt}
\caption{\label{BetaSingDet} Detailed view of the energy distributions of the 
``from beam'' and 
``off beam'' single track events in the vicinity of the electron end-point energy for a defined 
range of electron emission angles. The agreement between the measured (solid line) and 
simulated signal distribution (full circles) is near to perfect.}
\vspace*{-5pt}
\end{figure}

Such a comparison is shown in Fig.\,\ref{BetaSing} for single tracks and for the Mott 
scattering events. 
In the latter case the modification of the $\beta$ spectrum induced by the energy 
and  angular dependence of the Mott-scattering cross section is clearly visible.
Electronic thresholds are not included in the simulation --- this is why the measured and 
simulated distributions do not match at the low energy side. 
In contrast, the matching at the high energy side is nearly perfect for single track events 
(Fig.~\ref{BetaSingDet}).
Due to not well defined energy losses (in particular the determination of the depth at which 
the Mott scattering took place within the lead scatterer is far beyond the accuracy of the 
electron tracking) the analogous comparison, at a similar level of accuracy, is not possible 
in the case of V-track events.

\begin{figure}[htb] 
\includegraphics[scale=.44]{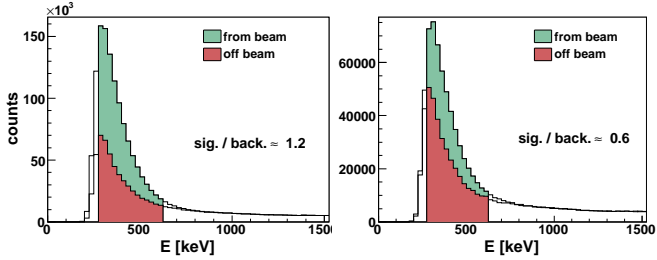}
\caption{\label{En_From_Out_1ver}(Color online) Energy distribution of signal (green) and 
background contributions (red) for single vertex events in anodes and cathodes (left and 
right, respectively). Shaded areas indicate the ``signal'' energy range used in the final calculations.}
\vspace*{0pt}
\end{figure}

The lack of two important matching conditions is the reason of the worse signal to background ratio 
observed in event classes with the single vertex signature (Fig.~\ref{En_From_Out_1ver}). 
The difference between events with the vertex reconstructed either in cathodes or in anodes 
can be explained by the worse double track resolution of cathodes and the horizontal geometry of the beam.

\begin{figure}[htb] 
\includegraphics[scale=.44]{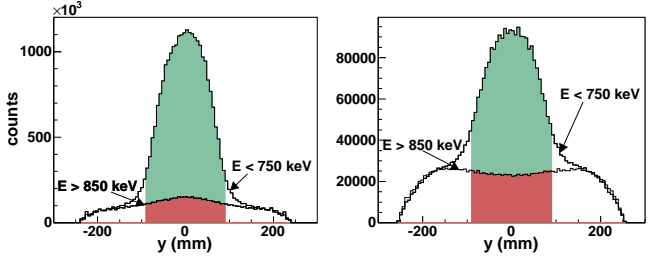}
\caption{\label{PosAtBeam}(Color online) Beam profile seen in the position distribution of 
the extrapolated electron origins for single track (left) and  V-track events (right), for 
low and high electron energies. 
Shaded areas indicate the position range accepted as ``from beam'' region.}
\end{figure}

The same background type can be observed in the  $y$-projection of the position 
distributions of the extrapolated origins for low and high energy electrons. 
The origin of the electron track was taken to be the intersection between its trajectory and the 
symmetry plane of the detector ($x=0$). 
This approximation is justified by the narrow beam size along the $x$-coordinate 
($\pm~$2~cm) defined by the $^{6}$Li beam collimator. 
The clear profile of the neutron beam can be recognized in the position distributions of low energy 
electrons ($E<750$~keV) in contrast to the distributions of events with higher energy 
(Fig.~\ref{PosAtBeam}).
The significant difference in the background distribution for double vertex events as compared 
to the case of single 
track events can be explained by the tighter setting of the geometrical cuts and by the 
dependence of the V-track detection efficiency on the Mott scattering angle. 
As already discussed, for more acute vertices the probability that clusters overlap increases, 
what as a consequence enhances the contribution of events originating in the off-the-beam volume with 
more obtuse vertices.

\begin{figure}[htb]
\includegraphics[scale=.44]{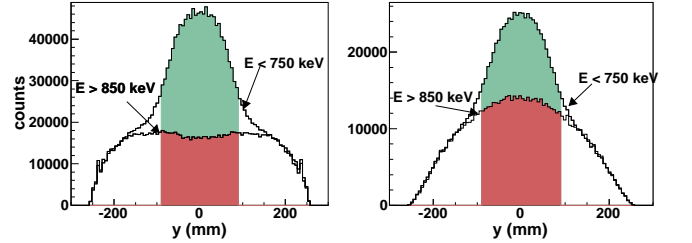}
\caption{\label{PosAtBeamV}(Color online)  Beam profile seen in the position distribution of 
the extrapolated electron origins for events with a single vertex in anodes (left) and 
cathodes (right), for low and high electron energies. Shaded areas indicate the position range 
accepted as ``from beam'' region.}
\end{figure}

The above argument does not apply to events with a single vertex in cathodes (Fig.~\ref{PosAtBeamV}). 
In this case the shape of the background resembles that of single tracks but the signal to 
background ratio is much worse than for events for which only one vertex was reconstructed in 
anodes. 
This can be explained by the beam fiducial volume geometry which is well defined in the 
vertical direction ($\pm90$~mm) but with horizontal limits fixed only as a compromise 
between striving to enhance statistics and to minimize the background of electrons originating 
from the MWPC frames. 

\begin{figure}[htb]
\includegraphics[scale=.44]{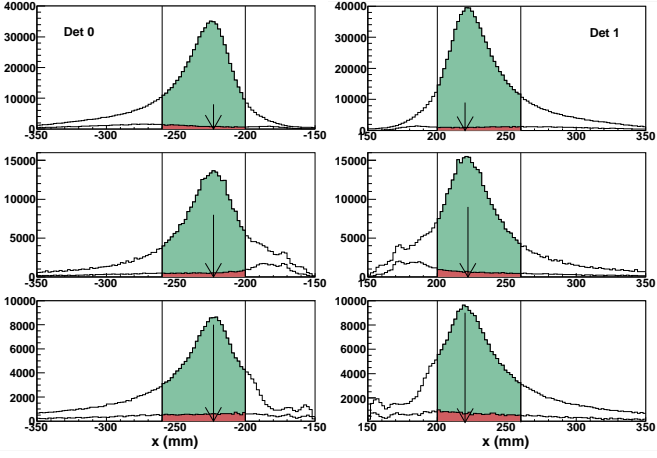}
\caption{\label{FinFoutMF}(Color online) ''Foil-out'' background contributions (red) to the 
vertex $x$-coordinate position distributions of V-track events, for the left and right detector 
side (left and right panels, respectively), separately for double-vertex events (top panels) 
and events with only one vertex (reconstructed in anodes -- middle, and cathodes -- bottom 
panels).  The arrows show the Mott foil position. Shaded areas indicate the position range 
accepted as ``from-foil'' region.}
\end{figure}

In the case of V-track events, beside the background discussed above, 
events for which the backscattering took place in the surrounding of the 
Mott target induce an additional source of background. 
Figure \ref{FinFoutMF} presents the distributions of the reconstructed $x$-component of the 
vertex positions for the data collected with and without Mott target.
The ``foil-out'' distributions have been scaled appropriately  by a factor 
deduced from the accumulated neutron beam for each setting of the Mott scatterer.

The ``foil-in'' distributions clearly peak at the Mott target position.
The broad maxima observed in ``foil-out'' distributions can be explained by backscattering 
on the MWPC material (mainly on the Ni-Cr wires and the aluminized exit window) and on the 
wrapping of scintillator hodoscopes.
The very good signal to background ratio observed for the double vertex events ($\approx$ 23) 
decreases for single vertex events in anodes ($\approx$ 16) to reach its minimum for events 
with a single vertex in cathodes ($\approx$ 9). 
The sharp structures seen in the distributions of one-vertex events, for $|x|<200$ are due to an 
artifact of the reconstruction procedure caused by the assignment of single wire clusters 
to the discrete wire positions. 
\subsection{\label{S4_6}  Beta decay asymmetry}

\begin{figure}[htb]
\hspace*{-4mm}
\includegraphics[scale=.44]{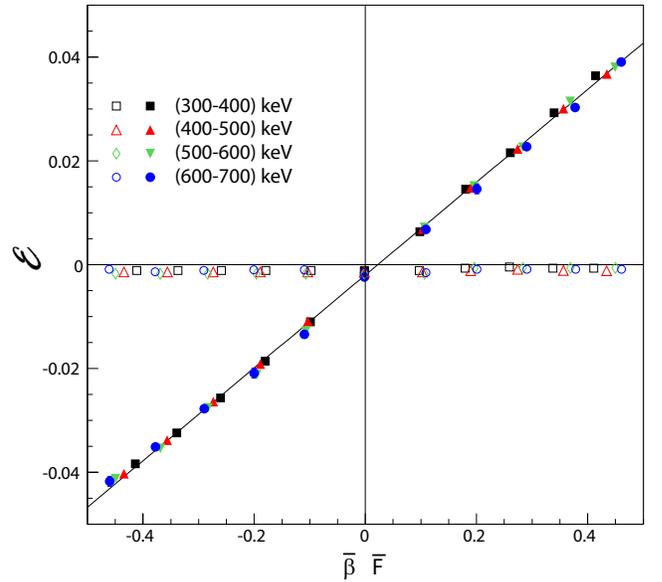} 
\caption{\label{Polar}(Color online)  Asymmetry $\mathcal{E}$ (Eq.~\ref{ASY3}) of signal 
(full symbols) and background (open symbols) as a function of 
$\overline{\beta\, F}$ for different electron energy ranges for events registered on 
one detector side. The fit of Eq.~(\ref{Polar}) to the data allows to extract the neutron beam 
polarization. A tiny negative shift of all $\mathcal{E}$ values is due to the spin flipper 
influence on the  detection efficiency $\epsilon^{+}$. }
\end{figure}

To extract the beam polarization $P$  averaged over the beam fiducial volume  the following 
asymmetries were analyzed:
 \begin{equation}\label{ASY1}
   \mathcal{E} \left(\beta, \gamma \right) = 
\frac
{N^{+}\!\left(\beta, \gamma \right) -N^{-}\!\left(\beta, \gamma \right)} 
{N^{+}\!\left(\beta, \gamma \right) +N^{-}\!\left(\beta, \gamma \right)},
 \end{equation}
where $N^{\pm}$ are experimental, background corrected numbers of counts of single tracks, 
sorted in 4 bins of the electron velocity normalized to the speed of light $\beta=v/c$, and 11 
bins of the electron emission angle  $\gamma$ with respect to the  neutron polarization direction.
Considering only the relevant terms in Eq.~(\ref{Wprob2}),  $N^{\pm}$ can be written as:
 \begin{equation}\label{ASY2}
N^{\pm}\left(\beta, \gamma \right) =
  N_{0}\,\epsilon^{\pm}[ 1 + \eta^{\pm}\, A\, P\cdot{\overline {\beta \,\mathcal F(\beta, \gamma)}}] ,
 \end{equation}
where the sign in superscripts reflects the beam polarization direction and $\beta \mathcal{ F}$ 
is a kinematical factor corresponding to the average $z$-component of the electron 
velocity in a given bin of $\beta$ and $\gamma$: 
 \begin{equation}\label{ASY2a}
\overline{ \beta \mathcal{ F}(\beta, \gamma)} = \,\langle \frac{{\bf v_e}}{c}\cdot {\bf \hat{J}}\rangle_{\beta, \gamma}
 \end{equation}
with ${\bf \hat{J}}$ a unit vector in the direction of the neutron polarization.
The factors $\eta^{\pm}$ account for the spin flipper efficiency.
When the spin flipper is switched off one has  $\eta_{-}\!\equiv\!1$ (original polarization 
is fully maintained), whereas in the opposite case $\eta_{+}=-1 +\eta$. 
The value  
$\eta\approx0.0114$, indicates that only a very small fraction of all neutrons had not 
reversed their spin after passage through the  spin flipper generated RF field. 
The magnitude of this effect has been investigated in a dedicated experiment \cite{Zej04}.
The factors $\epsilon^{\pm}$ account for the influence of the spin flipper operation on the 
detection efficiency of single track events. 
The radio frequency associated with the spin flipper operation, propagating via electrical 
grounding and in the air via the beam line volume, increases slightly the noise level 
observed on the wires of MWPCs and in the hodoscopes. 
This increases the dead time and decreases the reconstruction efficiency.
Similarly, as in the previous case $\epsilon^{-}\!\equiv\!1$ and $\epsilon^{+}=1-\epsilon$ is 
close to, but less than unity. 
The actual value of $\epsilon$ ($ \approx 0.004$) can be calculated from the total numbers of 
single track events 
accumulated in each beam polarization state and corrected for the corresponding beam intensity. 
It is visible as a small negative offset of all experimental points in Fig.~\ref{Polar}.
With  the above definitions, applying first order Taylor expansion in  $\epsilon$, 
$\eta$ and second order in the term $ P\,A\,\overline {\beta\,\mathcal F(\beta,\gamma)}$, 
Eq.~(\ref{ASY1}) reads:
 \begin{equation}\label{ASY3}
   \mathcal{E} \left(\beta, \gamma \right) = 
(1\!-\eta/2) P\, A \,\overline{ \beta\,\mathcal F(\beta, \gamma)} -\epsilon/2.
 \end{equation}
Figure \ref{Polar} shows the obtained  $\mathcal{E}$ as a function of 
$\overline{\beta\, F}$ for different electron energy ranges. 
Taking $A$ as a constant known with a very good precision ($A=-0.1173\pm0.0013$ \cite{PDG10}), 
the average neutron polarization can be obtained from a one-parameter fit of Eq.~(\ref{ASY3}) 
to the experimental data.

It should be noted that in the background correction procedure special care has been taken 
to ensure that the background counts do not depend on the beam polarization direction. 
The $\mathcal{E}$ asymmetries for events which do not originate from the beam volume 
(Fig.~\ref{Polar}, open symbols) as well as of the high energy events (above the electron end 
point energy) do 
not depend on the angle $\gamma$ and are consistent with zero polarization of their sources.

The average neutron polarization values for the four data taking periods are listed  in 
Table \ref{tab:table2}.
The low polarization for the 2004 data set has been traced to a bug in the guiding field found 
{\em post factum} and verified in a dedicated experiment.

The electron emission asymmetry should also be observed  in Mott scattered event classes.
Due to the much lower statistics of those events, the extracted polarization $P_{V}$ is much less 
precise. 
However, a similar analysis has been performed and its results are in satisfactory 
agreement with the single track data (Table~\ref{tab:table2}). 

\begin{table}[htb]
\vspace*{-12pt}
\caption{\label{tab:table2}Summary of neutron beam polarization analysis. Polarizations deduced 
from double and single vertex event classes ($P_{VV}$ and $P_{V}$) are also shown.}
\begin{ruledtabular}
\begin{tabular}{lccc}
Run &  $P\times 10^{2}$  & $P_{VV} \times 10^{2}$ & $P_{V} \times 10^{2}$    \\
\hline
2003     &    80.3$\pm$1.3$\pm$1.6  & 71.8$\pm$9.4$\pm$1.6 &   \\
2004     &    44.2$\pm$0.4$\pm$1.5  & 48.7$\pm$8.3$\pm$1.5 &   \\
2006     &    80.0$\pm$1.0$\pm$1.5  & 82.9$\pm$3.9$\pm$1.5 & 74.1$\pm$9.5$\pm$1.5  \\
2007     &    77.4$\pm$0.2$\pm$0.7  & 78.7$\pm$1.7$\pm$1.2 & 79.9$\pm$3.0$\pm$1.2  \\
\end{tabular}
\end{ruledtabular}
\end{table}
For the longest data taking period in 2007, an independent analysis of the beam polarization has 
been performed.
This approach used the same asymmetries as defined in Eq.~(\ref{ASY1}), however, with different 
binning.
The single track events were sorted in only 2 bins in $\gamma$ (electron 
emission into lower and upper hemispheres) but, additionally, in 2787 time bins (half an hour 
long each).
This allowed to search for a possible time dependence of the extracted polarization and daily 
modulations of different observables, and also provided a consistency check with the previous 
analysis.
\begin{figure}[htb]
\includegraphics[scale=.44]{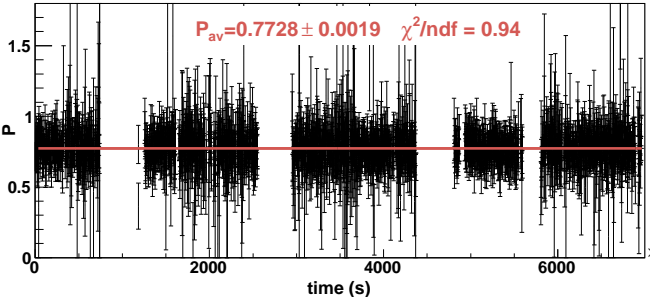}
\caption{\label{Pol_TimeSer} Time series of polarization extracted from the analysis of single 
tracks sorted in 2787 time bins. The longer breaks correspond to no-beam periods.}
\end{figure}
The constant value fit to the data (Fig.~\ref{Pol_TimeSer}) results in an average 
polarization of 0.773 $\pm$ 0.002 which is in a very good agreement with the result of the 
previous method. 
Considering only the statistical errors, the $\chi^{2}$ per degree of freedom of this fit amounts 
to 0.94. 
The discrete Fourier transform of the obtained time series is consistent with a white noise 
distribution. 
One can conclude that no significant time dependence has been observed within about 3 months of 
the 2007 data collection period. 
For more details  of this analysis and obtained limitations on some Lorenz invariance violating 
parameters see Ref.~\cite{Koz10b}.

\subsection{\label{S4_7}  Correlation coefficients $R$, $N$ }

\begin{figure}[htb]
\includegraphics[scale=.84]{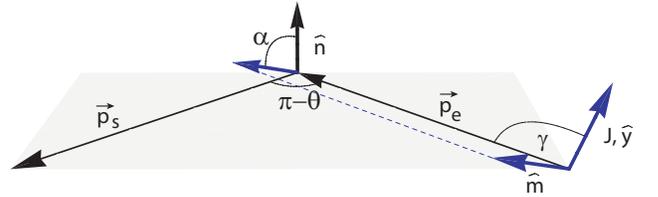}
\caption{\label{nTRV_Vectors_simpl} Definition of angles relevant for the analysis of the 
transverse electron polarization by Mott scattering. ${\bf J}$ is the neutron spin and 
${\bf p_{e}}$, ${\bf p_{s}}$ are incident and scattered electron momenta, respectively.}
\end{figure}

For the analysis of the transverse electron polarization components and the associated   
correlation coefficients, the method of backward Mott scattering was applied.
Parity and time reversal conservation of the spin-orbit force responsible for the spin 
dependence of this electromagnetic process guarantee its exclusive sensitivity to the 
transverse polarization component perpendicular to the scattering plane. 
Technically, this can be expressed by the following substitution:

\begin{equation}\label{HatSigma}
{\bf\hat{\sigma}} \rightarrow S(E, \theta) \,{\bf\hat{n}},
 \end{equation}
where $S$ is the effective analyzing power of the Mott-scatterer which for a given target 
element depends on electron energy $E$ and scattering angle $\theta$, and ${\bf\hat{n}}$ is 
a unit vector perpendicular to the Mott-scattering plane (Fig.~\ref{nTRV_Vectors_simpl})
\begin{equation}\label{HatN}
{\bf\hat{n}} = \frac{{\bf p_{e}}\times{\bf p_{s}}}{|{\bf p_{e}}\times{\bf p_{s}}|},
 \end{equation}
where ${\bf p_{e}}$ and ${\bf p_{s}}$ are incident and scattered electron momenta, respectively. 

\begin{figure}[htb]
\hspace*{-2mm}
\includegraphics[scale=.44]{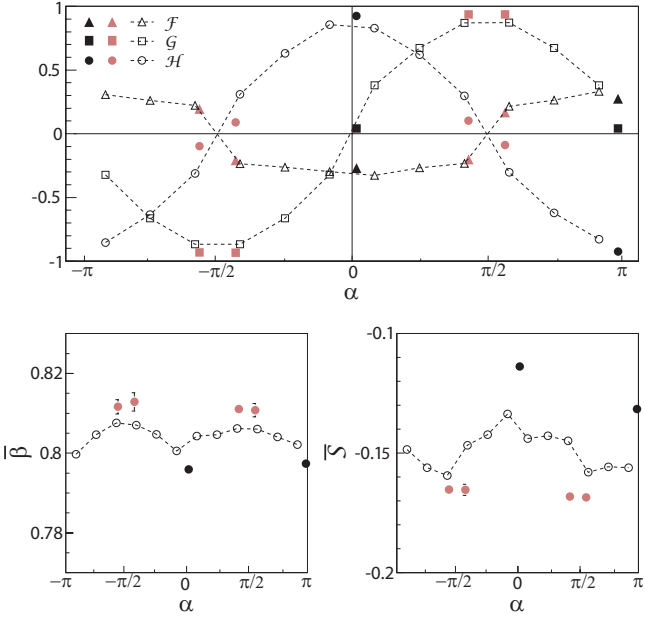}
\caption{\label{FormFacts}(Color online) Average geometrical ($\bar{\mathcal{F}}$, $\bar{\mathcal{G}}$, 
$\bar{\mathcal{H}}$) and kinematical  ($\bar{\mathcal{\beta}}$, $\bar{\mathcal{S}}$) 
factors as a function of $\alpha$ for double (open symbols) and single vertex events (full 
black and red symbols refer to vertical and horizontal V-tracks, respectively).  
Dotted lines are to guide the eye only. Error bars are smaller than the symbol size.}
\end{figure}

Applying this substitution to Eq.~(\ref{Wprob2}), the background-corrected 
experimental numbers of counts of V-track events $n^{\pm}$ can be expressed as:
 \begin{multline}\label{NRASY2}
 n^{\pm} = n_{0}\,\, \epsilon^{\pm}\!\!_{_{V}}\, \bigg\{1 +
 \eta^{\pm}P \Big[ A\,\overline{\beta\, \mathcal{F}(\alpha)}  + \\
 N  \overline { S \mathcal{G}(\alpha)} + 
R \, \overline{\beta S \mathcal{H(\alpha)}} \Big]\bigg\}.
 \end{multline}
The sign in superscripts reflects the beam polarization direction, the meaning of  
$\epsilon^{\pm}\!\!_{_{V}}$ and $\eta^{\pm}$ is the same as in the case of single track 
events, Eq.~(\ref{ASY2}), 
and the kinematic factors $ \mathcal{\bar{F}}(\alpha)$, 
$\mathcal{\bar{G}}(\alpha)$ and 
$ \mathcal{\bar{H}}(\alpha)$  represent the average values of the quantities 
${\bf \hat{J}\cdot\hat{p}}$,   ${\bf \hat{J}}\cdot\mbox{\boldmath$ \hat{\sigma}$}$ and
${\bf \hat{J}\cdot\hat{p}}\times\mbox{\boldmath$ \hat{\sigma}$}$, respectively 
(Fig.~\ref{FormFacts}). 
The bar over a term indicates event-by-event averaging used in all analyzing methods applied in 
this work.
In order to fully exploit the symmetry properties of both the physical problem and the 
experimental setup, all those quantities where sorted in 12 bins of  $\alpha$, defined 
as the angle between the electron-scattering and neutron-decay planes 
(Fig.~\ref{nTRV_Vectors_simpl}), represented respectively by the ${\bf\hat{n}}$ and  
${\bf\hat{m}}$  unit vectors.
\begin{equation}\label{HatM}
{\bf\hat{m}} = \frac{{\bf J}\times{\bf p_{e}}}{|{\bf J}\times{\bf p_{e}}|}.
 \end{equation}

\begin{figure}[htb]
\hspace*{-3mm}
\includegraphics[scale=.45]{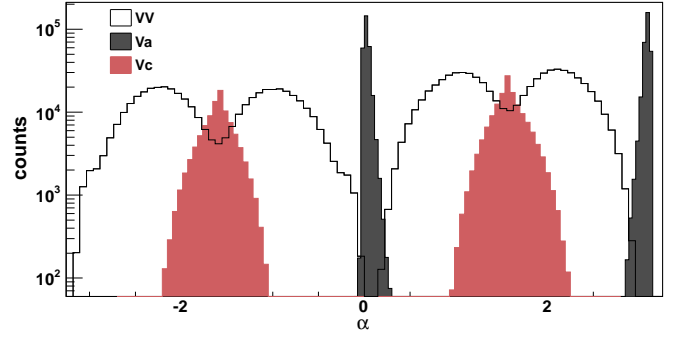}
\caption{\label{AlfaAng}(Color online) The $\alpha$  distribution ($n^{+}+n^{-}$) of all 
analyzed double vertex events ($V\!V$) and events with only one vertex reconstructed in 
anodes ($V_{a}$), and cathodes ($V_{c}$). }
\end{figure}

The $\alpha$ angle distribution of the sum $n^{+}+n^{-}$ for all classes of the 
analyzed V-tracks is shown in Fig.~\ref{AlfaAng}.
The deep minima around integer multiplicities of $\pi/2$,  in the distribution of double 
vertex events (VV) reflect the rectangular geometry of the MWPC and are mainly shaped by 
limited double track resolution of anodes and cathodes. 
The significantly lower  intensity at negative values of $\alpha$ angle (corresponding to concave 
V-tracks) is due to the magnetic field influence on the V-track detection efficiency, 
discussed in Section~\ref{S4_4}.
 
The visible spread of the  $n^{+}+n^{-}$ distribution for ``vertical'' single vertex events ($V_{a}$) 
is entirely an effect of the energy dependent correction for the magnetic field applied to 
the electron trajectories. 
Without this correction only two discrete values (0 and $\pi$)  would be possible 
(all relevant vectors ${\bf p_{e}}$, ${\bf p_{s}}$ and ${\bf J}$ are coplanar in this case).

Two different approaches have been used to obtain the $N$ and $R$ correlation coefficients. 
The first one, presented in the next section, can be applied to both V-track event classes, 
those with full geometrical information and those with only one vertex in the vertical plane. 
The results of this approach have been adopted as the final result of this experiment.
The second approach allows the extraction of  $N$ and $R$ coefficients separately from dedicated 
double ratios. 
However, in the case of the $R$ correlation this approach requires the assumption that the 
experimental data are symmetric with respect to the transformation $\alpha \rightarrow 
\alpha' = -\alpha$. 
This requirement is drastically violated by the influence of the spin holding magnetic 
field in the case of the ``vertical'' V-track event class. 
In the case of events with full geometrical information the $\alpha \rightarrow -\alpha$ 
symmetry is much better fulfilled, allowing for the final application of the double ratio 
method, albeit, as a consistency check only. 

\subsubsection{\label{S4_7_1}  Correlation coefficients R, N from asymmetry}

To extract the $N$ and $R$ correlation coefficients the following set of 
asymmetries was considered:
 \begin{equation}\label{NRASY1}
  \mathcal{A}\left(\alpha \right) = 
\frac
{n^{+}\left(\alpha \right) - n^{-}\left(\alpha \right)} 
{n^{+}\left(\alpha \right) + n^{-}\left(\alpha \right)}
 \end{equation}
Applying  Eq.~(\ref{NRASY2}) and the first order Taylor expansion in small quantities 
($\eta$, $\epsilon\!_{_{V}}$, $R\overline{S\beta \mathcal{H}}$ and  
$N \overline{S  \mathcal{G}}$) and second order Taylor expansion in the largest term, 
$P A \overline{\beta \mathcal{F}}$, one obtains: 
\begin{multline}\label{ASY4}
 \mathcal{A}\left(\alpha \right) = P (1-\eta/2)\,\Big[A \overline{\beta \mathcal{F}(\alpha)}\,+\\
	N \overline{S(\alpha) \mathcal{G}(\alpha)} + 
	 R \overline{S(\alpha) \beta  \mathcal{H}(\alpha)} \Big] - \epsilon\!_{_{V}}/2,
 \end{multline}
\begin{figure}[hbt]
\includegraphics[scale=.44]{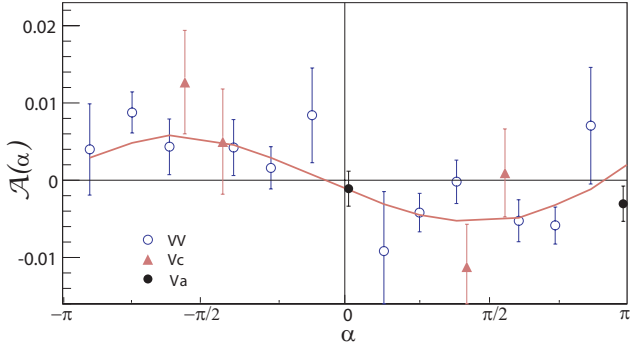}
\caption{\label{SNGRH} (Color online) Experimental asymmetries $ \mathcal{A}$ corrected for the  
$P A \overline{\beta \mathcal{F}}$ term  as a function of $\alpha$. 
The solid line illustrates the two-parameter ($N$, $R$) least-square fit to the data using 
experimental form factors  $\overline{\mathcal{S G}}$ and $\overline{\beta \mathcal{S H}}$. 
The indicated errors are of statistical nature.
}
\end{figure} 
The term $P A \overline{\beta \mathcal{F}}$ accounts for the nonuniform illumination of the 
Mott foil due to the $\beta$-decay asymmetry and is known precisely from event-by-event averaging.
The systematic uncertainty of this term is dominated by the error of the average 
beam polarization $P$.  
It is interesting to note that the functions  $\bar{\mathcal{G}}$ and $\bar{\mathcal{H}}$
follow quite closely sine and cosine functions, respectively, and are almost orthogonal to 
each other.
As a consequence, the covariance matrix of the two-parameter fit used to obtain 
the $N$ and $R$ correlation coefficients is almost diagonal, with the correlation coefficient 
$\rho(R,N)\approx$0.007.

A two parameter fit of the experimental asymmetries $ \mathcal{A}$, corrected for the 
 $P A \overline{\beta \mathcal{F}}$ term, to the experimental data set of 2007 is 
shown in Fig.\,\ref{SNGRH}.
The extracted values for the $R$ and $N$ coefficients are listed in 
Table \ref{tab:table3}.

\begin{table*}[htb]
\caption{\label{tab:table3}Summary of results obtained in all data collection periods. 
Statistical and systematic uncertainties follow the experimental values. 
$N_{S\!M}$  is the SM value of the $N$ coefficient calculated at 
$\bar{E}_{K}$. 
Its error comes from the experimental uncertainty of the decay asymmetry parameter $A$ \cite{PDG10}. 
The difference as compared to the results presented in \cite{Koz09} is due to an improved 
determination of the effective analyzing powers and including the additional class of single 
vertex events ($V$). $\chi^{2}/n.d.f.$ of all fits is also presented. }
\begin{ruledtabular}
\begin{tabular}{ccccccccccc}
Run & Ev. Class
         & $N_{S\!M}\!\times\!10^{3}$ 
                    &  $N\!\times\!10^{3}$
                                       & $R\times\!10^{3}$
                                                            & $\chi^{2}/n.d.f.$
                                                                  &$N\!\times\!10^{3}$ 
                                                                                        & $\chi^{2}/n.d.f.$
                                                                                             &$R\times\!10^{3}$ & $\chi^{2}/n.d.f.$\\
    &    &          & Eq.~(\ref{ASY4}) & Eq.~(\ref{ASY4})   & Eq.~(\ref{ASY4})
                                                                  & Eq.~(\ref{SR2})    &Eq.~(\ref{SR2})
                                                                                             & Eq.~(\ref{SR4})    &Eq.~(\ref{SR4})\\
\hline
2003& VV & 71$\pm$1 & 89$\pm$92$\pm$31 & -90$\pm$137$\pm$38 & 1.6 & 139$\pm$124$\pm$27 & 1.9 & -55$\pm$152$\pm$42 & 1.4\\
2004& VV & 68$\pm$1 & 74$\pm$80$\pm$17 & -135$\pm$130$\pm$30& 1.8 & 171$\pm$103$\pm$15 & 2.0 & -58$\pm$148$\pm$30 & 1.5\\
2006& VV & 68$\pm$1 & 94$\pm$35$\pm$10 & -13$\pm$48$\pm$10  & 1.3 &  97$\pm$35$\pm$10  & 0.6 & -36$\pm$48$\pm$12  & 2.1\\
2006& V  & 68$\pm$1 & 44$\pm$109$\pm$23& -50$\pm$55$\pm$21  &     & 53$\pm$117$\pm$23  &     &                    &    \\
2007& VV & 68$\pm$1 & 59$\pm$13$\pm$5  & 13$\pm$18$\pm$6    & 1.1 &  63$\pm$14$\pm$5   & 1.3 & -5$\pm$18$\pm$6    & 0.7\\
2007& V  & 68$\pm$1 & 51$\pm$32$\pm$14 &  9$\pm$20$\pm$13   &     &  52$\pm$33$\pm$5   &     &                    &    \\
\hline                                                                                          
Total&   &          & 62$\pm$12$\pm$4  &  4$\pm$12$\pm$5    &     &  67$\pm$11$\pm$4   &     &   &                \\
\end{tabular}
\end{ruledtabular}
\vspace*{-8pt}
\end{table*}

\subsubsection{\label{S4_7_2}  Correlation coefficients R, N from double ratios}

\begin{figure}[hbt]
\includegraphics[scale=.42]{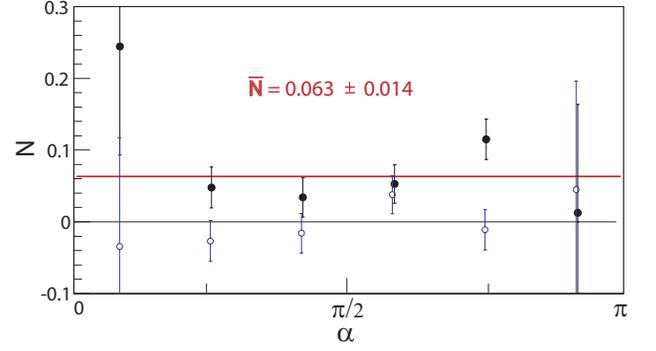}
\caption{\label{N_super} $N$ correlation coefficients calculated according to Eq.~(\ref{SR2}) 
for double vertex events (full symbols) as a function of  $\alpha$ and their average.  
The indicated errors are of statistical nature.  
Open symbols: the same but for unpolarized beam (see Sec.~\ref{S4_7_3}).}
\end{figure}

From the approximate symmetry of the detector with respect to the transformation 
$\alpha \rightarrow - \alpha$, it follows that  $ \bar{\beta}$, $\bar{S}$ and 
the factors $ \mathcal{\bar{F}}$, $ \mathcal{\bar{H}}$ are almost symmetric 
while $ \mathcal{\bar{G}}$ is an almost antisymmetric function of $\alpha$ (Fig.\,\ref{FormFacts}). 
Applying these symmetries, the Taylor expansion as in the previous section and the definition 
of the quantities $n^{\pm}$, Eq.~(\ref{NRASY2}), one can see that the double ratio defined as:
\begin{equation}\label{SR1}
 Q(\alpha) =  \frac{(r(\alpha)\!- \!1)}{(r(\alpha)\! +\!1)},
\end{equation}
where
\begin{equation}\label{SR1a}
    r(\alpha) = \sqrt{\frac{n^{+}(\alpha) \, n^{-}(-\alpha )} 
		{n^{-}(\alpha )\,  n^{+}(-\alpha )} },
\end{equation}
allows to extract the $N$ correlation coefficient according to:
\begin{equation}\label{SR2}
 N \approx  Q\cdot \frac{1-\frac{1}{2}\big((1-\eta/2)P A \overline{\beta F}\big)^2}
{(1-\eta/2) P \overline{S\, \mathcal{G}}}.
\end{equation}
The advantage of this method is that the effect associated with the 
term $P A \overline{\beta  \mathcal{F}}$ is suppressed by a factor of about 60 as 
compared to Eq.~(\ref{ASY4}).
The ratio $Q$ is also insensitive to the spin flipper related modulation of the 
detection efficiency.
Figure \ref{N_super} shows the values of $N$ obtained as a function of the angle $\alpha$ with 
their average value. 
The good agreement between the $N$ values obtained in both ways (Table \ref{tab:table3}) 
enhances our confidence in the experimental values of the $N$ and $R$ coefficients obtained 
in the previous section.

An alternative way to extract the $R$ correlation coefficient makes use of the 
analysis of another ratio:
\begin{equation}\label{SR3}
 U(\alpha) =  \frac{r'(\alpha)\!- \!1}{r'(\alpha)\! +\!1},
\end{equation}
where:
\begin{equation}\label{SR3a}
    r'(\alpha) = \sqrt{\frac{n^{-}(\alpha) \, n^{-}(-\alpha )} 
		{n^{+}(\alpha )\,  n^{+}(-\alpha )} }
\end{equation}
Applying Eq.~(\ref{NRASY2}) and keeping only terms linear in small quantities 
($\epsilon_{\!_{V}}$ , $P A \overline{\beta F}$, $P R \overline{\beta S\,\mathcal{G}}$), 
one can show that:
\begin{equation}\label{SR4}
 R \approx   \frac{U-(1-\eta/2)P A \overline{\beta F} - \epsilon_{\!_{V}}/2}
{(1-\eta/2) P\overline{\beta S\, \mathcal{H}}}.
\end{equation}
In this method one suppresses the term proportional to the $N$ correlation. 
It is, however, sensitive to the ``false'' asymmetry due to the term  $P A \overline{\beta 
\mathcal{F}}$ and to the spin flipper related modulation of the detection efficiency, and 
therefore has no clear advantage over the method based on Eq.~(\ref{ASY4}).
 \begin{figure}[hbt]
\includegraphics[scale=.42]{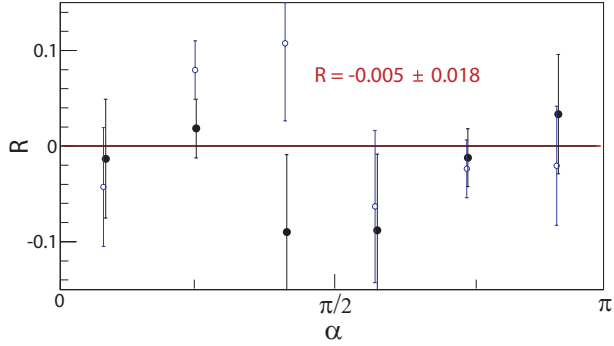}
\caption{\label{R_super} $R$ correlation coefficient calculated according to Eq.~(\ref{SR4}) 
for each  $\alpha$ bin for double vertex events.  Only statistical errors are indicated.}
\end{figure}
Moreover, since the distribution of events with only one vertex in anodes is not symmetric 
with respect to the transformation $\alpha \rightarrow - \alpha$ (Fig.~\ref{AlfaAng}), 
this approach can not be applied to this event class. 
Figure \ref{R_super} shows the obtained values of $R$ as a function of the angle $\alpha$ with 
their average. 
The  results for the double vertex event class are shown in Fig.~\ref{R_super} as a function 
of $\alpha$ with their average, and are also included in Table \ref{tab:table3}. 

\subsubsection{\label{S4_7_3}  Polarization of background and unpolarized beam}

\begin{figure}[hbt]
\includegraphics[scale=.42]{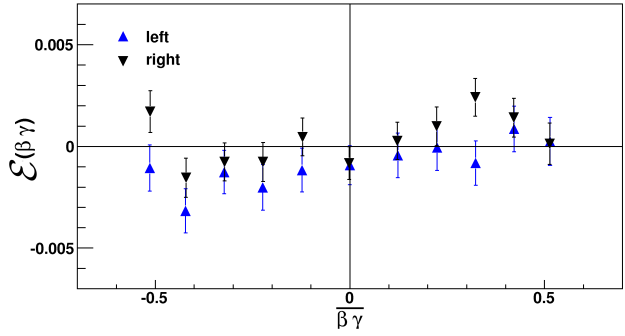}
\caption{\label{BetAsy_HEbackgr} Asymmetry  $\mathcal{E} \left(\beta, \gamma \right)$ 
for single-track events with energy larger than the neutron $\beta$-decay end-point energy, 
separately for the left and right detector side.}
\end{figure}

\begin{figure}[hbt]
\includegraphics[scale=.44]{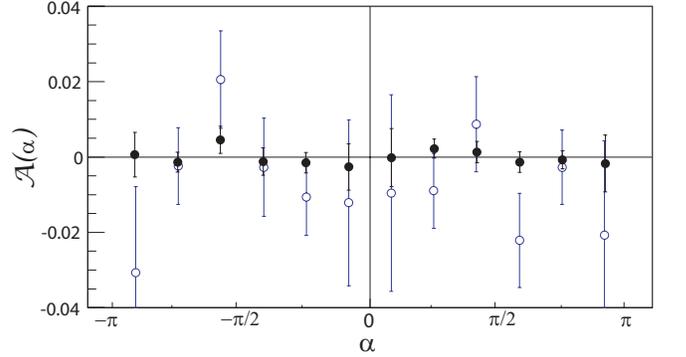}
\caption{\label{AsyHe} Asymmetry  $\mathcal{A} (\alpha )$ 
for double-vertex events with energy larger than the neutron $\beta$-decay end-point energy
(open symbols), and for unpolarized beam (full symbols).}
\end{figure}

An important consistency check of the analysis of the Mott scattered events relied on 
the determination of the background polarization. 
As there was no conceivable mechanism which could cause such polarization the expected 
value was zero.
To check this presumption, the asymmetries as defined in Eqs.~(\ref{ASY1}) and (\ref{NRASY1}) 
have been calculated for events with energy larger than the neutron $\beta$-decay end-point 
energy (Figs.~\ref{BetAsy_HEbackgr} and \ref{AsyHe}) and for events originating outside 
the beam fiducial volume (Fig.~\ref{Polar}). 
\begin{figure}[hbt] 
\includegraphics[scale=.44]{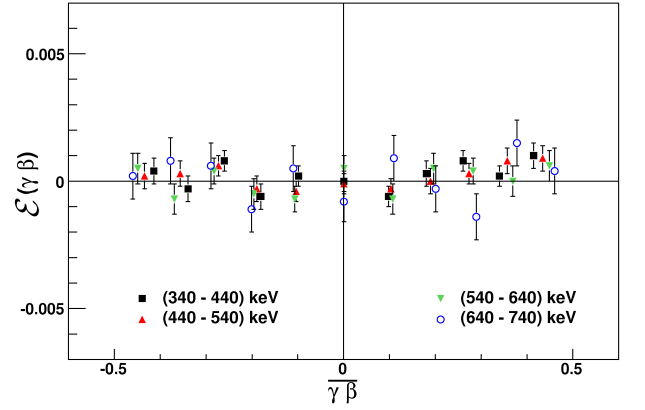}
\caption{\label{BetAsy_BCDE_Unpol} Asymmetry  $\mathcal{E} \left(\beta, \gamma \right)$ 
for single-track events and unpolarized beam. Both detector sides have been added together. }
\end{figure} 
It turned out that within the statistical accuracy all asymmetries were  consistent 
with zero.
This proves that the analysis was not biased by, for instance, a spin flipper related false 
asymmetry.

The same test has been performed for the data taken with an unpolarized beam. 
In this case, the analysis differed from the regular analysis of polarized beam in only one 
detail: each two consecutive spin states corresponding to one flipper ``on'' and one 
flipper ``off'' periods were concatenated into a new one called ``spin-up'', while the next 
two were used to obtain a new ``spin-down'' state.
This construction assures averaging of the polarization over an equal number of neutrons in 
both spin states and hence leads to an unpolarized beam (we neglect here the effect due to spin 
flipper inefficiency which is well below the statistical accuracy of this analysis). 

Also in this case the asymmetries were consistent with zero (Fig.~\ref{BetAsy_BCDE_Unpol}) what 
strengthens the confidence in the obtained final results. 

\subsection{\label{SystErr} Systematic errors}

The systematic uncertainties involved in the evaluation of the $R$ and $N$ coefficients are 
dominated by effects introduced by the background subtraction procedure and the 
choice of specific values of the cuts which determine whether an individual event is 
attributed to the ``signal'' or to the ``background''.  

\begin{figure}[hbt]
\includegraphics[scale=.44]{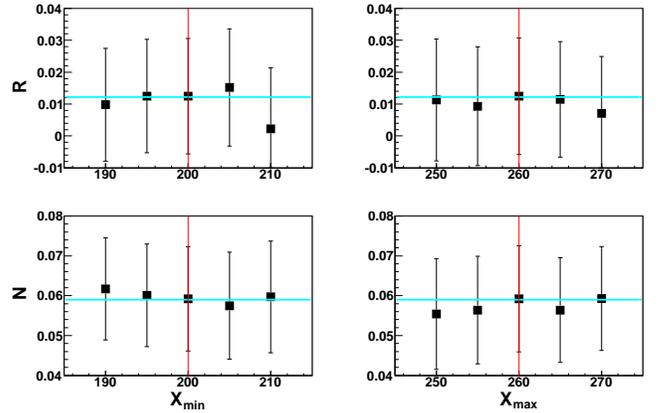}
\caption{\label{NorRSyst_FromFl07OK} Influence of the geometrical definition of the accepted 
scattering vertex positions on the reconstructed $R$ and $N$ correlation coefficients. 
Vertical lines indicate values applied in the analysis. 
}
\end{figure}

A ``signal'' event is defined as an electron originating from the free neutron decay 
and backscattered off the Mott target.
From this definition it follows that the set of necessary conditions describing such event 
must include:
(i) geometrical limitations to the beam volume, (ii) specification of the allowed energy 
range and (iii)  geometrical limitation to the area of the Mott target applied to the 
reconstructed scattering vertex.
As a general rule, the symmetry of the detector setup has been preserved in the definition 
of cuts.
The single exception from this rule was made for the beam limitation along the $y$ 
coordinate: the cutting line was inclined by an angle of about 1$^{o}$ in accordance with the 
beam divergence.
This allowed limiting the number of parameters to the following three groups (see also 
Fig.~\ref{setup}):
\begin{itemize}
\item
``from/off beam'' definition: $y_{1_{max}}$,  $z_{1_{max}}$,  $y_{2_{max}}$, 
\item
from/off neutron decay: $E_{L_{min}}$,  $E_{L_{max}}$, $E_{H_{min}}$,  $E_{H_{max}}$,
\item
from Mott foil: $X_{min}$,  $X_{max}$,  $W_{max}$ (side length of the square indicating the 
foil area in the $y\!-\!z$ plane).
\end{itemize}
Those parameters were used to classify an event as belonging to:
\begin{enumerate}
\item
 signal:
\begin{eqnarray}\label{CondSign}
|y| \,<& y_{1_{max}},\quad |z|\,<& z_{1_{max}},\\
\nonumber E \,>& E_{L_{min}}, \quad E \,<& E_{L_{max}},\\
\nonumber X_{min} > X >  X_{max}, &\quad  |Y|<  W_{max},& \quad  |Z|<  W_{max},
\end{eqnarray}
\item
off beam background: as above but 
\begin{eqnarray}\label{CondBack1}
|y| \,> y_{2_{max}} 
\end{eqnarray}
\item
high energy background: 
\begin{eqnarray}\label{CondBack2}
|y| \,<& y_{1_{max}},\quad |z|\,<& z_{1_{max}},\\
\nonumber E \,>& E_{H_{min}}, \quad E \,<& E_{H_{max}},\\
\nonumber X_{min} > X >  X_{max}, &\quad  |Y|<  W_{max},& \quad  |Z|<  W_{max},
\end{eqnarray}
\end{enumerate}
where $(x\!=\!0,y,z)$ and $(X,Y,Z)$ denote coordinates of the electron origin and 
the Mott scattering vertex, respectively. 
Due to the limited accuracy of the reconstructed energies and trajectories, for each of 
those parameters there exists a certain range of values which seems to be almost equivalent.
In the analysis, however, the specific values can generate slightly different final results.  
In order to estimate this effect, the corresponding ranges of acceptance have been identified 
individually for each parameter and the final analysis was repeated varying one parameter in 
its range with all others fixed in the center of their ranges. 
The maximal deviations of the resulting $R$ and $N$ coefficients from the central values were 
taken as maximal errors.
Figure \ref{NorRSyst_FromFl07OK} presents the result of such analysis for parameters limiting the 
geometrical position of the Mott scattering vertices. 
The contributions of all parameters to the final uncertainty, as well as their ranges used in this 
analysis, are collected in Table \ref{tab:table1}. 

It should be noted that the vertical alignment of the apparatus with respect to the neutron beam 
has been verified to a precision below 1~mm using the reconstructed centroid of the beam 
profile (Figs.~\ref{PosAtBeam} and \ref{PosAtBeamV}).
A similar precision of alignment with respect to the magnet setup and the beam axis was 
maintained  for all other detector components except the Mott scattering target. 
Here the accuracy of positioning in the $x$-coordinate was about $\pm$2~mm. 
This is still acceptable considering the vertex reconstruction accuracy of the MWPCs 
(Fig.~\ref{FinFoutMF}). 

The next important systematic uncertainty  is due to the limited accuracy of the 
determination of the average beam polarization. 
All measured asymmetries used for the evaluation of individual correlation 
coefficients are proportional to the product of this coefficient and the beam 
polarization.
As a consequence, the relative error of the extracted coefficient must be larger than that 
of the polarization.

The situation is even more difficult for the $R$ correlation. 
With a vertically polarized neutron beam the existence of a nonzero value of this 
correlation would result in a difference between the number of electrons backscattered 
into the upper and into the lower hemisphere.
The same effect is generated by the $\beta$-decay asymmetry ($A$ correlation), appearing as a 
term $P A \overline{\beta \mathcal{F}}$ in Eqs.~(\ref{ASY4}) and (\ref{SR4}). 
What makes the link between the $A$ and $R$ correlations so special is the 
shape of the corresponding average form factors $\mathcal{\bar{F}}$ and $\mathcal{\bar{H}}$ 
(Fig.~\ref{FormFacts}).
They both exhibit the same symmetry properties with respect to $\alpha$, so  
that the effects generated by each of them are almost indistinguishable.

\begin{figure}[hbt]
\includegraphics[scale=.43]{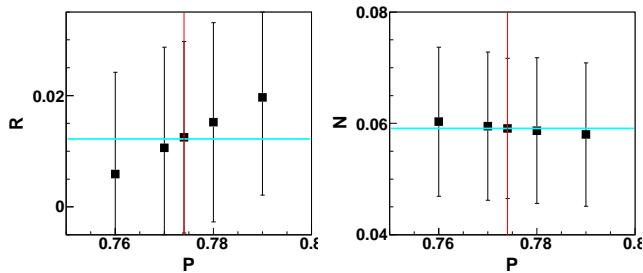}
\caption{\label{RNSyst_Pol07OK} Dependence of the reconstructed $R$ and $N$ correlation 
coefficients on the value of the neutron beam polarization. Vertical lines indicate values 
applied in the analysis. The strong correlation between $R$ and $P$ is due to the  
$P\beta A\mathcal{\bar{F}}$ term in Eq.~(\ref{ASY4}).} 
\end{figure}

The cleanest, but also unpractical, way to avoid this interference would be to confine 
the electron emission angle to  90$^{o}$, in which case  $\mathcal{\bar{F}} \equiv 0$. 
With a finite accepted solid angle one is forced to apply a suitable correction in the form
of the $P A \overline{\beta \mathcal{F}}$ term. 
Since the electron momenta after emission and after scattering are reconstructed, the 
necessary form-factors ($\mathcal{\bar{F}}$, $\mathcal{\bar{G}}$, $\mathcal{\bar{H}}$) 
are known to a high precision for the entire event sample, and the main 
impact of the applied correction is due to the uncertainty of the neutron polarization. 
To evaluate the magnitude of the influence of this term on the final result, 
the fit with $R$ and $N$ as free parameters was repeated with $P$ varied by one 
standard deviation of the total neutron polarization uncertainty (Fig.~\ref{RNSyst_Pol07OK}). 
The obtained difference enters the budget of the systematic errors and is presented in 
Table \ref{tab:table1}. 

Despite the careful design and manufacturing of the large volume electromagnet responsible 
for the spin holding magnetic field, the mapping of this field at the experimental position 
showed a small misalignment with respect to the vertical direction and nonuniformities in 
the beam fiducial volume. 
The average effect was accounted for in the analysis by appropriate rotation of the neutron 
polarization direction. 
A residual systematic effect (Table  \ref{tab:table1}) was induced by the uncertainty of 
the field measurements and by the observed stability of the ambient magnetic field at the 
experimental position.

Since the radio--frequency signal of the spin flipper was a source of small noise in the 
readout electronics, tiny dead time variations  correlated with the spin flipper were 
observed. 
Their influence on the result was corrected for. 
The residual effect is included in Table  \ref{tab:table1}.

\begin{table}[htb]
\vspace*{-12pt}
\caption{\label{tab:table1}Summary of systematic errors for the 2007 data set. 
The ranges over which individual parameters have been varied in the error estimation procedure 
are shown.}
\begin{ruledtabular}
\begin{tabular}{lccc}
Source &  $\delta N \times 10^{4}$  & $\delta R \times 10^{4}$ \\
\hline
 $y_{1_{max}} \in (80,95)$       & 20 &  17   \\
 $z_{1_{max}} \in (240,250)$     & 16 &   9   \\
 $y_{2_{max}} \in (100,120)$     & 26 &  25   \\
 $E_{L_{min}} \in (200,260)$     & 13 &  11   \\
 $E_{L_{max}} \in (580,660)$     & 18 &  15   \\
 $E_{H_{min}} \in (780,860)$     &  5 &   4   \\
 $E_{H_{max}} \in (1400,1900)$   &  7 &  16   \\
 $X_{min}  \in (190,210)$        & 11 &   18  \\
 $X_{max}  \in (245,265)$        & 15 &   18  \\
 $W_{max}  \in (210,230)$        & 19 &   23  \\
term $P A \overline{\beta \mathcal{F}}$ & 6  & 30 \\
effective Sherman function $\bar{S}$      & 13 & 4  \\ 
guiding field misalignment       & 3  & 7  \\
dead time variations             & 9  & 0.5\\
\hline
Total                            & 54 & 61 \\
\end{tabular}
\end{ruledtabular}
\end{table}

In the final error analysis it has been assumed that the sources contributed independently, 
and so these were added quadratically to obtain the final systematic uncertainty.

\section{\label{S5}  Results}

Combining the results from all runs leads to the final result (Table \ref{tab:table3}):
\begin{align}
R=0.004\pm0.012_{\text{stat}}\pm0.005_{\text{syst}},\\
N=0.067\pm0.011_{\text{stat}}\pm0.004_{\text{syst}}.
\end{align}

\begin{figure}[hbt]
\includegraphics[scale=.61]{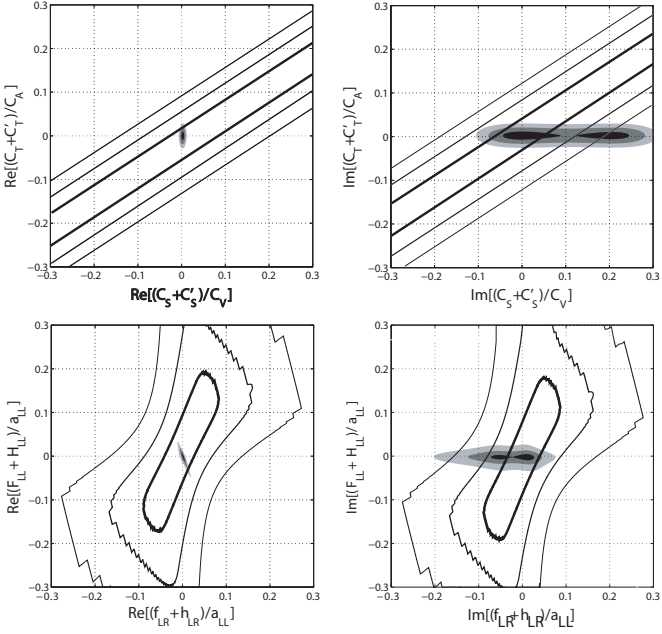}
\caption{\label{KB_rezultaty}Experimental bounds on the scalar vs. tensor normalized 
couplings (upper) and leptoquark exchange helicity projection amplitudes (lower panels). 
The grey areas represent the information as defined in
Ref. \cite{Sever06}, while the lines represent the limits resulting from the present experiment.
Decreasing line thickness as well as intensity of the grey areas correspond to 1-, 
2- and 3-sigma confidence levels.}
\vspace*{-14pt}\label{MSSMRPV}

\end{figure}

\begin{figure}[hbt]
\includegraphics[scale=.8]{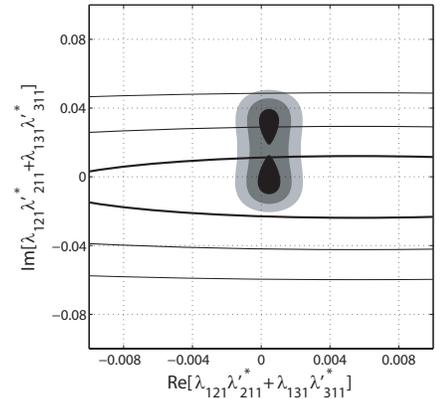}
\caption{\label{ExclPlotsRPV}Experimental bounds on the real vs. imaginary combined couplings 
of MSSM with R-parity violation. 
The grey areas represent the information as defined in Ref. \cite{Sever06}, while the lines 
represent the limits resulting from the present experiment.
Decreasing line thickness and intensity of the grey areas correspond to 1-, 
2- and 3- sigma confidence levels.}
\vspace*{-14pt}
\end{figure}

In Figs.~\ref{KB_rezultaty} and \ref{ExclPlotsRPV} the new results are presented as  
exclusion plots containing in addition the experimental information available to date
from nuclear and neutron beta decays, as surveyed in Ref.~\cite{Sever06}. 
The upper part of Fig.~\ref{KB_rezultaty} contains plots corresponding to the real and 
imaginary parts of the normalized scalar and tensor coupling constants $S$ and $T$, 
Eqs.~(\ref{DefS}) and (\ref{DefT}). 
The present accuracy of the determination of the $N$ correlation coefficient does not improve 
the already strong constraints on the real part of the couplings (left panel). 
It is, however,  consistent with the existing data and, in addition, adds confidence to the 
validity of the extraction of the $R$ correlation coefficient.
The latter constrains significantly  
the imaginary part of the scalar couplings, beyond the limits from all previous measurements 
(right panel).
Also in this case the result is consistent with the SM ($C_{S}=C'_{S}=C_{T}=C'_{T}=0$) and 
with time-reversal invariance.

In the lower part of Fig.~\ref{KB_rezultaty} the same convention was used to illustrate 
the constraints (existing and resulting from the present work) to the helicity projection 
amplitudes in the leptoquark exchange model, as defined in Ref. \cite{Herc01}.
In this formalism $F_{LL}, f_{LR}$ and $H_{LL}, h_{LR}$ correspond to leptoquarks with charge 
$|Q|=2/3$ and  $|Q|=1/3$, respectively.
Capital letters correspond to the scalar (spin-zero) while  lower-case letters describe vector 
(spin-one) leptoquarks exchange amplitudes. 
Subscripts indicate the helicity structure of the underlying interaction.  
As in the previous case, only the $R$ correlation reveals evident exclusion power and allows 
a significant improvement of the constraints on the imaginary part of the vector leptoquark 
amplitudes ($f_{LR} +h_{LR}$). 

Similar constraints can be imposed on the selectron exchange couplings ($\lambda_{1i1}$, 
$\lambda_{i11}$) in the minimal supersymmetric standard model (MSSM) with R-parity 
violation. 
Adopting conventions used in Ref.~\cite{Yam10}, the amplitude of the selectron 
($\tilde{e}$) exchange between the quark and lepton can be written as:
\begin{equation}y
 \mathcal{M}_{\tilde{e}_L} = \sum_{i=2,3} \frac{\lambda_{1i1}\lambda_{i11}'^{*}}{4m^{2}_{\tilde{e}_{Li}}} \, \bar{u}(1+\gamma_{5})d \cdot  \bar{e}(1-\gamma_{5})\nu_{e},
\end{equation}
where $m_{\tilde{e}_{L}}$ is the slepton mass assumed to be equal to 100~GeV. 
According to Ref.~\cite{Yam10}, the  contribution of the selectron exchange  to the scalar 
coupling of the beta-decay and to the $R$ and $N$ correlations can be written as:
\begin{align} 
C_{S} =  g_{s}\,\,\sum_{i=2,3} \frac{\lambda_{1i1}\lambda_{i11}'^{*}}{4m^{2}_{\tilde{e}_{Li}}} ,
\end{align}
\begin{align} 
R=-\lambda \frac{2\,\sqrt{2} \, Im C_{S}}{G_{F}\,V_{u,d}\,g_{V}(1+3\lambda^{2})},\\
N=-\lambda \frac{2\,\sqrt{2} \, Re C_{S}}{G_{F}\,V_{u,d}\,g_{V}(1+3\lambda^{2})},
\end{align}
where $g_{s}$ is given by the neutron, proton and light quarks masses:
\begin{align}
g_{s}=\frac{M_{n}-M_{p}}{m_{u}-m_{d}} \approx 0.49\pm0.17,
\end{align}
$V_{ud}=0.97425(22)$ is the CKM matrix element, 
$\lambda= -1.2694(28)$  and 
$G_{F}=1.166364(5)\times10^{-5}$ GeV$^{-2}$ \cite{PDG10}.

The presently best direct constraint for the imaginary part of the scalar interaction 
obtained in the present experiment improves significantly the limits on the combination of 
coupling constants leading to the updated version of the exclusion plot presented in 
Ref.~\cite{Yam10} (Fig.~\ref{ExclPlotsRPV}).

The real part of this combination was also accessible in this experiment via the measurement 
of the $N$ correlation. 
The achieved accuracy, even if slightly better than that for $R$, can not compete with the much 
more precise data adopted from the compilation of superallowed Fermi nuclear beta decays. 

\section{\label{S6}  Conclusion}

The measurement of the transverse polarization components of electrons from the decay of free 
polarized neutrons has been carried out successfully. 
This was the first experimental determination of the $R$  correlation coefficient in 
neutron $\beta$-decay and, to our knowledge, also the first observation of the finite value 
of the $N$ correlation, an effect of the final state interaction in the neutron 
$\beta$-decay. 
The obtained results allowed a significant improvement in constraining the relative strength of 
exotic, scalar-type weak interaction and related parameters in standard model extensions with 
leptoquark exchange and in the MSSM with R-parity violation beyond the limits from all previous 
measurements.

The most important feature of the experimental setup which made this possible was
the ability to fully reconstruct momenta  of low energy electrons before and after the 
backward Mott scattering which served as the electron polarization analyzer.

Further, even substantial improvement in the statistical accuracy of the determination of 
$R$ and $N$ correlation coefficients can be achieved in an experiment based on this 
principle, provided that a substantial increase of the solid angle acceptance is attained.

\section{\label{S7}  Acknowledgements}
Special thanks are due to ETH Zurich for the continuous support and to Jerzy Sromicki for 
his leading role at the initial phase of this project. 
This work was supported in part by Polish Committee for Scientific Research
under the Grants 2P03B11122, N202 047037, by an Integrated Action Program Polonium 
(Contract No.\,05843UJ) and by the FWO Vlaanderen.
Part of the computational work was performed at ACK Cyfronet, Krak\'ow. 
The collaboration is grateful to PSI for excellent support and kind hospitality.
\newpage 
\bibliography{PRC_nTRV}

\end{document}